\documentclass[twocolumn]{aastex63}
\usepackage{ulem}
\received{July 11, 2022}
\revised{January 3, 2023}
\accepted{January 24, 2023}
\submitjournal{ApJ}

\shorttitle{Evidence for the Operation of the Hanle and Magneto-Optical Effects}
\shortauthors{Ishikawa et al.}

\begin{document}

\title{
Evidence for the Operation of the Hanle and Magneto-Optical Effects in the Scattering 
Polarization Signals Observed by CLASP2 Across the Mg~{\sc ii} $h$ \& $k$ Lines}

\correspondingauthor{Ryohko Ishikawa}
\email{ryoko.ishikawa@nao.ac.jp}

\author[0000-0001-8830-0769]{Ryohko Ishikawa}
\affiliation{National Astronomical Observatory of Japan, \\
2-21-1 Osawa, Mitaka, Tokyo 181-8588 Japan}

\author[0000-0001-5131-4139]{Javier Trujillo Bueno}
\affiliation{Instituto de Astrof\'{i}sica de Canarias, E-38205 La Laguna, Tenerife, Spain}
\affiliation{Departamento de Astrof\'{i}sica, Universidad de La Laguna, E-38206 La Laguna, Tenerife, Spain}
\affiliation{Consejo Superior de Investigaciones Cient\'{i}ficas, Spain}

\author[0000-0001-9095-9685]{Ernest Alsina Ballester}
\affiliation{Instituto de Astrof\'{i}sica de Canarias, E-38205 La Laguna, Tenerife, Spain}
\affiliation{Departamento de Astrof\'{i}sica, Universidad de La Laguna, E-38206 La Laguna, Tenerife, Spain}

\author[0000-0002-8775-0132]{Luca Belluzzi}
\affiliation{IRSOL Istituto Ricerche Solari ``Aldo e Cele Dacc\`{o}'', Universit\`{a} della Svizzera italiana (USI), \\
CH-6605 Locarno-Monti, Switzerland}
\affiliation{Leibniz-Institut f\"{u}r Sonnenphysik (KIS), D-79104 Freiburg, Germany}
\affiliation{Euler Institute, Universit\`{a} della Svizzera italiana (USI), CH-6900 Lugano, Switzerland}

\author[0000-0003-1465-5692]{Tanaus\'{u} del Pino Alem\'{a}n}
\affiliation{Instituto de Astrof\'{i}sica de Canarias, E-38205 La Laguna, Tenerife, Spain}
\affiliation{Departamento de Astrof\'{i}sica, Universidad de La Laguna, E-38206 La Laguna, Tenerife, Spain}

\author[0000-0002-9921-7757]{David E. McKenzie}
\affiliation{NASA Marshall Space Flight Center, Huntsville, AL 35812, USA}

\author[0000-0003-0972-7022]{Fr\'ed\'eric Auch\`ere}
\affiliation{Institut d'Astrophysique Spatiale, 91405 Orsay Cedex, France}

\author{Ken Kobayashi}
\affiliation{NASA Marshall Space Flight Center, Huntsville, AL 35812, USA}

\author[0000-0003-3765-1774]{Takenori J. Okamoto}
\affiliation{National Astronomical Observatory of Japan, \\
2-21-1 Osawa, Mitaka, Tokyo 181-8588 Japan}

\author[0000-0002-3770-009X]{Laurel A. Rachmeler}
\affiliation{National Oceanic and Atmospheric Administration, \\ 
National Centers for Environmental Information, Boulder, CO 80305, USA}

\author[0000-0003-3034-8406]{Donguk Song}
\affiliation{Korea Astronomy and Space Science Institute 776, \\
Daedeokdae-ro, Yuseong-gu, Daejeon 305-348, Republic of Korea}
\affiliation{National Astronomical Observatory of Japan, \\
2-21-1 Osawa, Mitaka, Tokyo 181-8588 Japan}



\begin{abstract}
Radiative transfer investigations of the solar Mg~{\sc ii} $h$ \& $k$ resonance lines around 280~nm 
showed that, while their circular polarization (Stokes $V$) signals arise from the Zeeman effect,     
the linear polarization profiles (Stokes $Q$ and $U$) are dominated by 
the scattering of anisotropic radiation and the Hanle and magneto-optical (MO) effects. 
Using the unprecedented observations 
of the Mg~{\sc ii} and Mn {\sc i} resonance lines
obtained by the Chromospheric LAyer Spectro-Polarimeter (CLASP2), here 
we investigate how the linear polarization signals at different wavelengths  
(i.e., at the center, and at the near and far wings of the $k$ line) 
vary with the longitudinal component of the magnetic field ($B_{L}$) 
at their approximate height of formation.
The $B_{L}$ is estimated from the $V$ signals in the
aforementioned spectral lines.
Particular attention is given to 
the following quantities that are expected to be influenced by the presence 
of magnetic fields through the Hanle and MO effects: 
the sign of the $U$ signals, the 
total linear polarization amplitude ($LP$) and 
its direction ($\chi$) with respect to a reference direction.
We find that at the center and near wings of the $k$ line, the behavior of   
these quantities is significantly different in the observed quiet and plage regions, 
and that both $LP$ and $\chi$ seem to depend on $B_{L}$.
These observational results are indicative of the operation of the Hanle effect 
at the center of the $k$ line and of the MO effects at the near wings of the $k$ line.
\end{abstract}


\section{Introduction}
\label{sec:intro}
The remote sensing inference of the magnetic fields that permeate 
the solar atmosphere is crucial to understand the energy transfer from 
the photosphere to the chromosphere and corona, and its dissipation. 
Over the last decades we have achieved increasingly better determinations 
of the photospheric magnetic fields \citep[e.g.,][]{2019LRSP...16....1B}.
However, there is a serious lack of empirical information 
on the chromosphere and 
the layers above, where the magnetic field 
dominates the dynamics, structuring, and heating of the plasma.

Over the last ten years, a series of theoretical investigations pointed out that the scattering of anisotropic radiation and the Hanle and Zeeman effects should produce measurable polarization signals in several ultraviolet (UV) spectral lines, such as hydrogen Lyman-$\alpha$ at 121.6~nm and Mg {\sc ii} $h$ \& $k$ around 280~nm, which encode information on the outer layers of the solar chromosphere
\citep[see the review by][]{2017SSRv..210..183T}.
The Hanle effect is the magnetically induced modification of the linear polarization produced by scattering processes (i.e., scattering polarization) at the center of a spectral line \citep{Landi2004}. 
Radiative transfer (RT) calculations in one-dimensional (1D) models of the solar atmosphere show  
that the observational signatures found at lines of sight away from the solar disk center are the depolarization and rotation of the plane of linear polarization
\citep{Trujillo2011,2016ApJ...830L..24D,2016ApJ...831L..15A}.
However, the line-center scattering polarization 
is also strongly influenced by the vertical stratification of the atmosphere and the 
lack of axial symmetry of the incident radiation field (i.e., the radiation illuminating each point within the medium). 
The former influence causes the amplitudes of the scattering polarization to be different for different atmospheric models \citep{2017SSRv..210..183T,2020ApJ...891...91D}. The latter can be due to the horizontal inhomogeneities of the solar atmosphere (temperature and density variations) and to the gradients 
of the plasma macroscopic velocity 
\citep{2011ApJ...743...12M,2015ApJ...803...65S,2016ApJ...826L..10S,2021ApJ...909..183J}. 
For this reason, 
inferring the chromospheric magnetic field via 
the Hanle effect is not straightforward.
The same applies to the scattering linear polarization signals in the wings of 
hydrogen Lyman-$\alpha$ and Mg {\sc ii} $h$ \& $k$. Via the 
magneto-optical (MO) terms of the Stokes-vector transfer equation 
\citep[see Eq. 6.85 in][]{Landi2004}
these wing polarization signals are sensitive to the presence of magnetic fields in 
their region of formation
\citep{2016ApJ...830L..24D,2016ApJ...831L..15A,2019ApJ...880...85A,2020ApJ...891...91D}.
These wing polarization signals are also affected by the above-mentioned vertical stratification and the lack of axial symmetry.

The theoretical studies mentioned above
led to the development of a series of sounding rocket experiments: the Chromospheric Lyman-Alpha Spectro-Polarimeter (hereafter CLASP1) and the Chromospheric LAyer Spectro-Polarimeter (CLASP2), 
whose aim was to demonstrate that UV spectropolarimetry is a suitable  technique to determine the magnetic field in the upper chromosphere and the transition region. 

In 2015, CLASP1 provided the first measurements of the linear polarization 
signals (i.e., the Stokes $Q$ and $U$ parameters)\footnote{In this paper, 
the reference direction for positive Stokes $Q$ is 
the parallel to the nearest solar limb.} produced 
by scattering processes in the
hydrogen Lyman-$\alpha$ line 
\citep{2017ApJ...839L..10K}. 
The Hanle effect operates at the spectral line center and for the hydrogen Lyman-$\alpha$ line
the Hanle critical magnetic field strength is 53 G. 
A detailed analysis of the intensity (Stokes $I$) and the fractional linear polarization ($Q/I$ and $U/I$) profiles observed by CLASP1 close to the limb \citep[see][]{2017ApJ...841...31I}
revealed that the $U/I$ signals in the wings of the
hydrogen Lyman-$\alpha$ line, where the Hanle effect does not operate, always
changed their sign when crossing the center of four  
bright structures whose photospheric magnetic flux was 
measured by the Solar Dynamic Observatory \citep[SDO,][]{2012SoPh..275....3P}.
These bright structures were also identified at the center of the Lyman-$\alpha$ intensity profiles.
However, at  
the Lyman-$\alpha$ line center, such a change of sign in $U/I$ was found only in an inter-network 
bright structure, where the photospheric magnetic flux was found to be relatively low,  
but not in the aforementioned network and enhanced network bright structures, 
where the photospheric magnetic flux was found to be larger.  
On the basis of the idealized polarization maps obtained by \cite{2012ASPC..456...59S} 
with and without magnetic fields in a very simple 3D model of the solar atmosphere \citep[see also Figure 7 of ][]{2017ApJ...841...31I},   
the above-mentioned curious behavior of 
the observed Lyman-$\alpha$ $U/I$ signals 
was taken as evidence for the operation of the Hanle effect at the center 
of the Lyman-$\alpha$ line \citep{2017ApJ...841...31I}. 
We note that it remains to be clarified why the $U/I$ wing signals 
always changed their sign when crossing the center of such bright structures 
irrespective of their magnetic flux given that,  
via the MO terms of the Stokes-vector transfer equation, the wings of the $Q/I$ and 
$U/I$ profiles of the hydrogen Lyman-$\alpha$ line are indeed 
sensitive to the presence of magnetic fields with strengths similar 
to those that produce the Hanle effect at 
the line center \citep[see][]{2019ApJ...880...85A}.

In 2019, CLASP2 provided the first ever spectrally 
and spatially resolved Stokes profiles observations
(intensity $I$, linear polarization $Q$ and $U$, and circular polarization $V$) 
across the Mg {\sc ii} $h$ \& $k$ lines, in an active region plage and in a quiet region near the solar limb. 
The Zeeman-induced Stokes $V$ circular polarization signals
were detected in several UV spectral lines, and in particular 
in the Mg~{\sc ii} $h$ \& $k$ lines and in the nearby Mn~{\sc i} resonance lines. 
The amplitude of these $V/I$ signals was significant in the observed active region plage and in the surrounding 
enhanced network. By applying the weak field approximation (WFA),
\citet{2021SciA....7.8406I} could determine  
the longitudinal component ($B_L$) of the magnetic field from the bottom to the top of the chromosphere,
both in the plage and in the surrounding enhanced network features. 

CLASP2 measured the linear polarization across the Mg {\sc ii} $h$ \& $k$ lines in both quiet Sun and plage targets.
The $Q/I$ and $U/I$ profiles are dominated 
by the scattering of anisotropic radiation in the Mg {\sc ii} $h$ \& $k$ lines, the modeling of which 
requires taking into account the joint action of partial frequency redistribution (PRD) 
and quantum interference between the upper $J-$levels of these lines \citep{Belluzzi2012ApJ}. 
The presence of a magnetic field produces measurable modifications of the 
scattering polarization over a wide wavelength range 
across the Mg {\sc ii} $h$ \& $k$ lines \citep{2016ApJ...830L..24D,2016ApJ...831L..15A}.
Via the Hanle effect, 
the $Q/I$ and $U/I$ line center signals of the $k$ line are sensitive to 
the strength, between approximately $5$ and $125$~G,\footnote{The Hanle critical field strength is 25 G for the $k$ line.} as well as to the orientation of the magnetic fields present in the upper chromosphere.
For stronger fields the Hanle effect is in its saturation regime and the line center signals are 
only sensitive to the orientation of the magnetic field. 
Moreover, the MO terms of the 
Stokes-vector transfer equation introduce magnetic sensitivity in the wings of the 
Mg {\sc ii} $h$ \& $k$ lines \citep{2016ApJ...831L..15A,2016ApJ...830L..24D}.
Magnetic fields as weak as those capable of producing the Hanle effect
in the $k$ line (see above) are sufficient to 
significantly affect the polarization wings of these lines via the MO effects.
The typical observational signatures of the MO effects are a depolarization of the $Q/I$ wing signals
and the appearance of $U/I$ wing signals.
Equivalently, they are characterized by a decrease of the linear polarization degree and a rotation of the plane of linear polarization
\citep[e.g., see Figure~4 of][]{2016ApJ...831L..15A}. 

By combining the information provided by the Zeeman-induced
circular polarization in the Mg {\sc ii} $h$ \& $k$ lines 
and the linear polarization resulting from
the joint action of scattering processes 
and the Hanle and the MO effects, we can 
aim at determining the magnetic 
field vector. To this end, it is first important to identify the observational signature of the
Hanle and MO effects in the CLASP2 data. In a recent paper we 
demonstrated that the $Q/I$ and $U/I$ patterns observed by CLASP2 across the Mg {\sc ii} $h$ \& $k$ lines
are indeed caused by the scattering of anisotropic radiation, 
with PRD and $J$-state interference playing a key role 
\citep{2022ApJ...936...67R}. 
In this paper, we aim at providing
observational evidence of the operation of the Hanle effect in the core of 
the $k$ line and of the MO
effects in the wings of the Mg {\sc ii} $h$ \& $k$ lines.   

\section{Observations and Data Reduction}
\begin{figure}[ht!]
\includegraphics[angle=90,width=\linewidth]{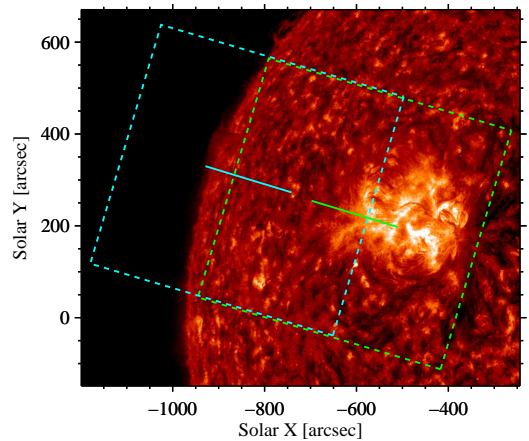}
\caption{SDO/AIA 30.4~nm intensity image temporally averaged 
between 16:53:41 and 16:58:41 UT on 2019 April 11.
Cyan and green boxes in dashed lines indicate the 
CLASP2/SJ fields of view
for the limb and plage targets, respectively.
The solid line at the center of each box indicates the CLASP2 spectrograph slit 
position for each target.
\label{fig:aia_img}}
\end{figure}

\begin{figure*}
\includegraphics[angle=90,width=\linewidth]{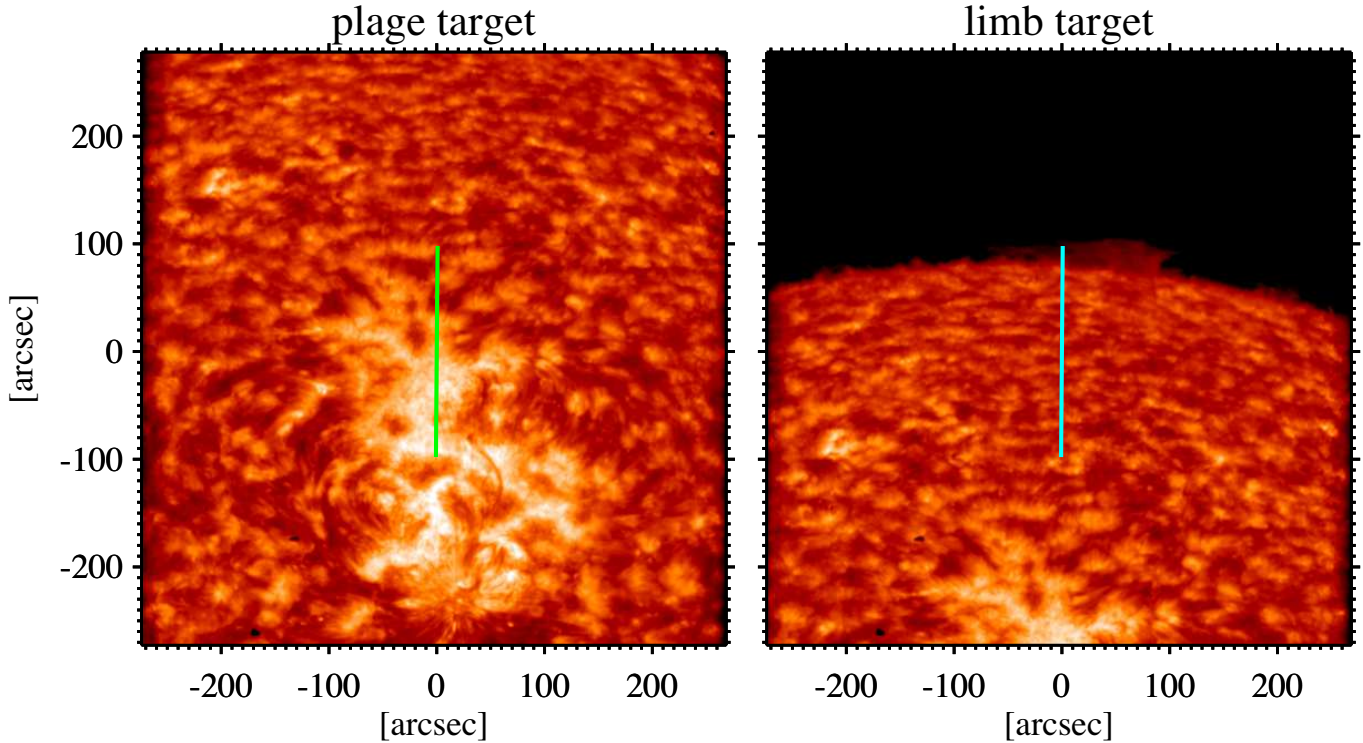}
\caption{Broadband (FWHM$=$3.5~nm) intensity images taken by the CLASP2/SJ at the
limb and plage targets. 
After correcting for pointing drift and jitter, the images were temporally averaged 
over 150.6~s and 137.8~s for the plage (left panel) and limb (right panel)
targets, respectively.
The CLASP2 spectrograph slit is indicated 
with solid lines following the color convention in Figure \ref{fig:aia_img}.
The horizontal and vertical axes give the distances from the center of the spectrograph slit.
\label{fig:sj_img}}
\end{figure*}

On 11 April 2019, CLASP2 conducted sit-and-stare observations 
at three positions on the solar disk.
The first target, observed during 15~s, was the solar disk center.
This observation allowed verifying that the instrumental polarization is negligible, roughly two orders of magnitude smaller than our required polarization accuracy of 0.1\% \citep{2022SoPh..297..135S}.
The second target was the East side of the NOAA 12738 active region 
(hereafter, plage target, see green box in Fig.~\ref{fig:aia_img})
observed from 16:53:40 to 16:56:16 UT (156~s).
Finally, it observed a quiet region near the limb 
(hereafter, limb target, see cyan box in Fig.~\ref{fig:aia_img})
from 16:56:25 to 16:58:46 UT (141~s).
The solid lines in Figure~\ref{fig:aia_img} showing the 30.4~nm image
obtained by the Atmospheric Imaging Assembly \citep[AIA;][]{2012SoPh..275...17L} on board the SDO, show the positions of the CLASP2 spectrograph slit, whose
length is $196\arcsec$ (Figure~\ref{fig:aia_img}). 
The slit was roughly oriented along the radial direction.

Figure~\ref{fig:sj_img} shows the broadband (full width at half maximum, $\mathrm{FWHM}=3.5$~nm) Lyman-$\alpha$ images 
obtained by the CLASP2 slit-jaw monitor system (CLASP2/SJ), including 
the exact positions 
of the slit. The CLASP2/SJ uses two Lyman-$\alpha$ broad-band filters
and it took chromospheric images  with a fast cadence of 0.6~s and a plate scale of 1\farcs03/pix \citep{Kubo2016}.
In the plage target (see the left panel of Figure~\ref{fig:sj_img}),
about one third of the slit covered the relatively quiet region 
outside the bright region of the plage. Concerning the limb target (see the right panel of Figure~\ref{fig:sj_img}),
about $\sim23\arcsec$ of the slit were outside of the Lyman-$\alpha$ limb.

The CLASP2 spectropolarimeter has two optically symmetric channels 
with their respective CCD cameras, which provide simultaneous measurements 
in two orthogonal polarization states \citep{2020SPIE11444E..6WT}.
The two cameras recorded the modulated radiation every 0.2~s in synchronization 
with a polarization modulation unit \citep[PMU:][]{2015SoPh..290.3081I} 
which continuously rotates a waveplate at 3.2~s/rot \citep{Ishikawa2013}.
The wavelength range covers 
from 279.3~nm to 280.7~nm 
including the Mg~{\sc ii} $h$ \& $k$ lines at 280.35~nm and 279.64~nm, respectively, 
and two Mn~{\sc i} lines at 280.19~nm and 279.91~nm (vacuum wavelengths).
After dark and gain corrections, we 
demodulated the observed signals   
to derive the wavelength variation of the Stokes $I$, $Q$, $U$, and $V$ parameters
for each PMU rotation and for each channel. Subsequently, 
we corrected for pointing and wavelength drifts and co-registered the $I$, $Q$, $U$ and $V$ spectra 
so that the slit and wavelength directions are aligned to the CCD pixels. 
Then, we applied the response matrix of the CLASP2 instrument to the $I$, $Q$, $U$, and $V$ spectra, 
which was derived in the polarization calibration investigation \citep{2022SoPh..297..135S}.
Finally, we obtain the $I$, $Q$, $U$, and $V$ spectra by combining the two channels.
The plate scales of the spectropolarimeter 
are 0.527\arcsec/pix along the slit and 0.00498~nm/pix along the dispersion direction.
The spatial and spectral resolutions are $\sim1.1\arcsec$ 
and $\sim0.01$~nm, respectively \citep{2018SPIE10699E..2WS,2018SPIE10699E..30Y}.
The reference direction for positive (negative) 
Stokes $Q$ is the parallel (perpendicular) to the nearest solar limb, while the reference direction for positive (negative) Stokes $U$ is at   
45$^\circ$ counterclockwise (clockwise) with respect to the $Q>0$ reference direction.

\begin{figure*}
\plotone{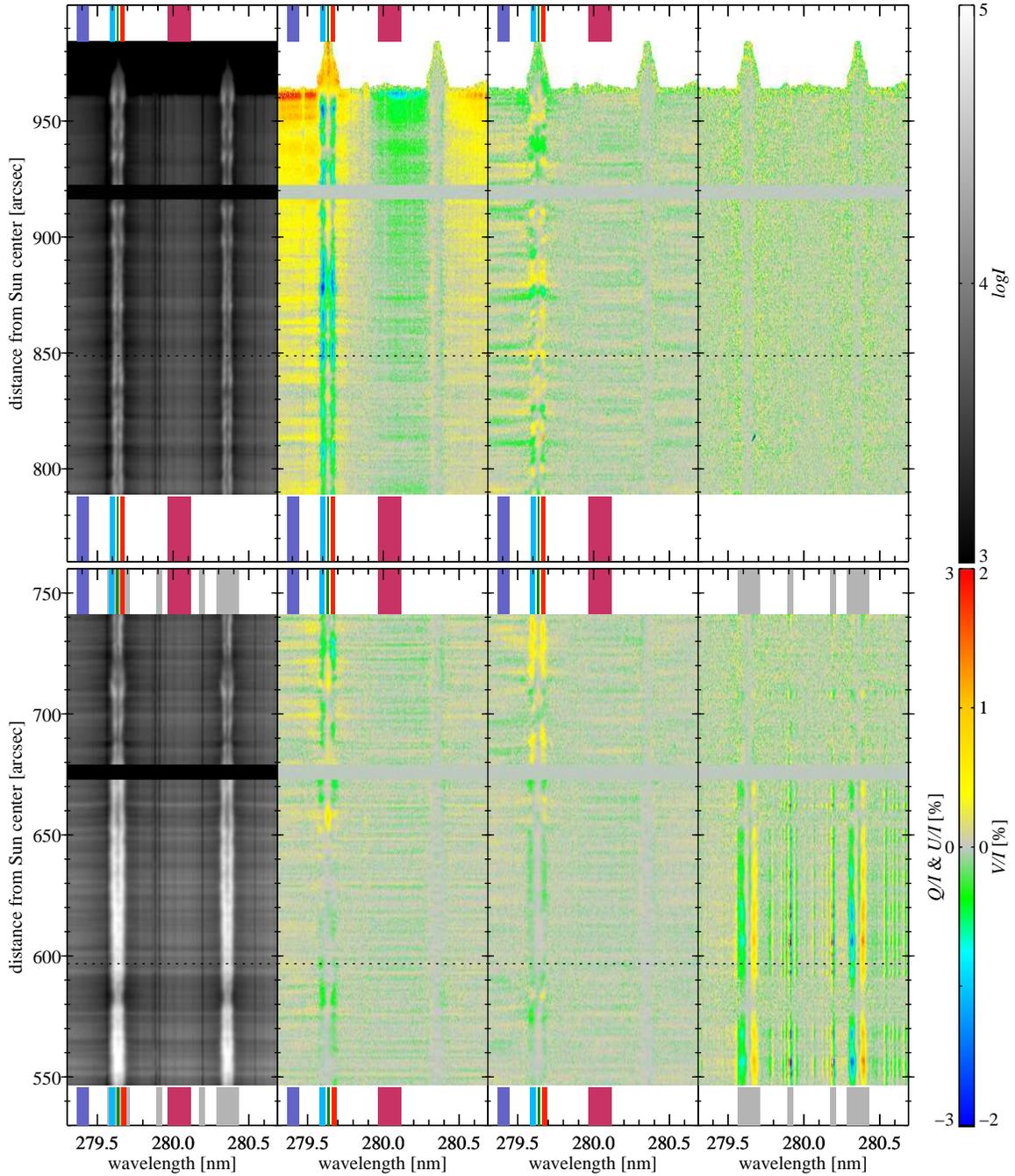}
\caption{From left to right, temporally averaged intensity $I$ and fractional polarization $Q/I$, $U/I$, and $V/I$ spectra across the Mg~{\sc ii}~$h$ \& $k$ lines obtained by CLASP2
at the limb and plage targets (upper and lower panels, respectively).
The intensity $I$ is the averaged number of electrons per exposure detected 
by the two channels in logarithmic scale. 
The data under the dust on the slit, which are shown in
black (in $I$) and gray (in $Q/I$, $U/I$, and $U/I$) horizontal stripes
are excluded in the analysis.
The Stokes signals corresponding to each of the 
spectral ranges colored with violet, light blue, green, red, and magenta have been   
integrated for obtaining the  
far blue wing, the near blue wing, the core, the near red wing, and the far red wing
signals explained in the text.
The colors are the same as those in Figures~\ref{fig:rga} and \ref{fig:rgb}.
The gray areas in the $I$ and $V/I$ panels indicate the spectral regions of the 
Mg~{\sc ii} $k$ (279.64~nm), 
Mn~{\sc i} blue (279.91~nm), Mn~{\sc i} red (280.19~nm), and Mg~{\sc ii} $h$ (280.35~nm) lines, respectively, across which we
have applied the WFA to estimate the longitudinal component of the magnetic field at different 
heights in the solar atmosphere.
\label{fig:iquv_spectra}}
\end{figure*}

In order to maximize the polarization  
sensitivity,
we use all the  
acquisitions during each pointing, resulting in temporally averaged Stokes spectra
of 150.4~s and 137.6~s for the plage and limb targets, respectively.
The time-averaging of the Stokes profiles smears out the influence of the dynamics \citep[e.g.,][]{2013ApJ...764...40C,2020ApJ...891...91D}.
The present investigation concerns the interpretation of the temporally-averaged Stokes profiles.
The final Stokes $I(\lambda)$ and fractional polarization spectra $Q(\lambda)/I(\lambda)$, $U(\lambda)/I(\lambda)$, and $V(\lambda)/I(\lambda)$, with $\lambda$ being the wavelength (hereafter, $I$, $Q/I$, $U/I$, and $V/I$), over the whole wavelength range 
obtained by CLASP2 are shown in Figure~\ref{fig:iquv_spectra}.
The distance from the disk center ranges from 546\farcs8 to 740\farcs9 ($0.64\le\mu\le0.82$) and from 789\farcs2 to 984\farcs0 ($\mu\le0.57$)
for the plage and limb targets, respectively.
$\mu$ is the cosine of the heliocentric angle, which was determined by taking the limb 
location at 961\farcs5 \citep{2022ApJ...936...67R}.

\begin{figure}
\plotone{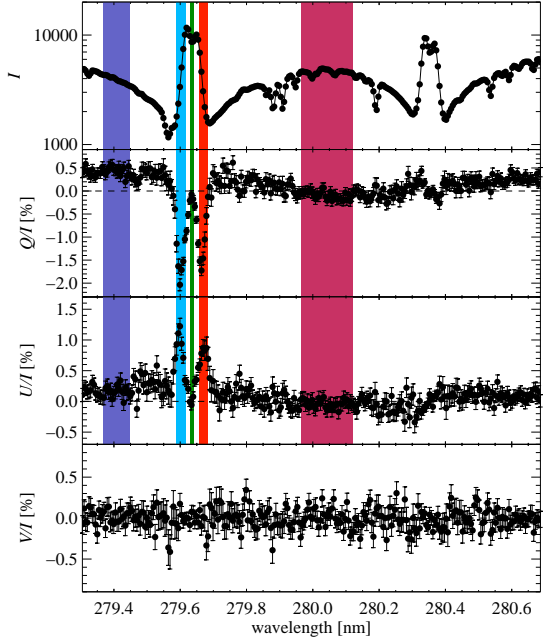}
\caption{Stokes profiles from the limb target at the
slit position indicated by the dotted line in the upper panel of Fig.~\ref{fig:iquv_spectra}. 
In the lower three panels, the error bars show $\pm1\sigma$ errors based on the photon noise.
The spectral regions 
colored with violet, magenta, light blue, red, and green 
are the same as those indicated in Fig.~\ref{fig:iquv_spectra}.
\label{fig:rga}}
\end{figure}

\begin{figure}
\plotone{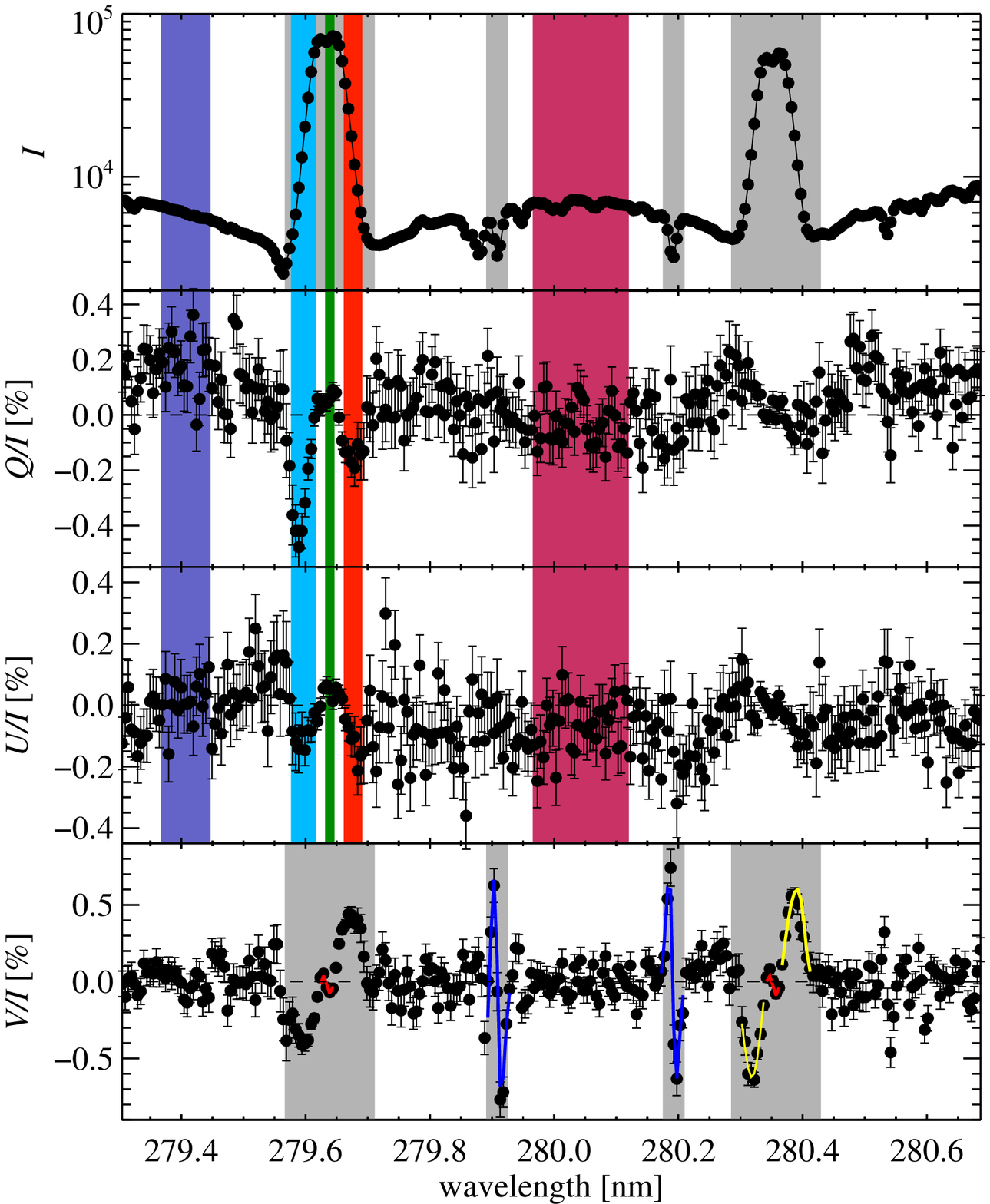}
\caption{Stokes profiles from the plage target 
at the slit position
indicated by the dotted line in the lower panel of Fig.~\ref{fig:iquv_spectra}.
In the lower three panels, the error bars show $\pm1\sigma$ errors based on the photon noise.
The spectral regions colored with violet, magenta, light blue, red, green, and gray 
are the same as those indicated in Fig.~\ref{fig:iquv_spectra}.
The red, yellow, and blue curves in the $V/I$ bottom panel  
show the fits that result from the application of the 
WFA to the inner lobes of the Mg~{\sc ii} $h$ \& $k$ lines, 
to the external lobes of the Mg~{\sc ii}~$h$ line, and to the Mn~{\sc i} lines, respectively.
\label{fig:rgb}
}
\end{figure}

\section{Data Analysis}
\subsection{Wavelength-integration of the linear polarization signals}\label{data_analysis_lp}

As clearly seen in Figure~\ref{fig:iquv_spectra},  
the Mg~{\sc ii} $k$ line at 279.64~nm 
shows significant fractional linear polarization $Q/I$ and $U/I$ signals,
while they are zero at the very center of the Mg~{\sc ii} $h$ line at 280.35~nm.
The $Q/I$ pattern observed by CLASP2 in a quiet region close to the limb
is consistent with the theoretical scattering polarization pattern 
predicted by \cite{Belluzzi2012ApJ} through radiative transfer calculations 
with partial frequency redistribution 
(PRD) and quantum-mechanical interference between 
the upper $J$-levels of the Mg~{\sc ii} $k$ \& $h$ lines.
Such calculations were carried out in 
the unmagnetized semi-empirical model C of \cite{Fontenla1993}, representative of the quiet solar atmosphere.
Recently, \cite{2022ApJ...936...67R}  
compared the observed center-to-limb variation (CLV) with such radiative transfer calculations. 
Figures~\ref{fig:rga} and \ref{fig:rgb} show
examples of the observed $I$, $Q/I$, $U/I$, and $V/I$ profiles from the quiet and plage targets, respectively.
 
In this paper, we investigate the linear polarization signals in five ranges 
in the observed spectra, specifically at 
the far blue and red wings of the $k$ line (violet and magenta in Figure~\ref{fig:iquv_spectra}, respectively), at 
the near blue and red wings of the $k$ line 
(i.e., between the $k_{2}$ and $k_{1}$ peaks, light blue and red in Figure~\ref{fig:iquv_spectra}, respectively),
and at the Mg~{\sc ii} $k$ core (i.e., 
a few pixels around the $k_{3}$ line center, green in Figure~\ref{fig:iquv_spectra}).
The Stokes signals at these wavelength ranges encode information on 
different layers of the solar atmosphere, roughly 
at the upper photosphere (far wings of the $k$ line), 
at the middle chromosphere (near wings of the $k$ line), 
and at the top of the chromosphere ($k$ core), respectively
\citep[e.g., see Figure 8 of][]{2020ApJ...891...91D}.
In order to characterize the behavior of the scattering polarization signals,  
we define the following five quantities,
\begin{equation}
i=\sum_{\lambda}I({\lambda}),
\label{eq:i}
\end{equation}
\begin{equation}
q=\frac{\sum_{\lambda}Q({\lambda})}{\sum_{\lambda}I({\lambda})},
\label{eq:q}
\end{equation}
\begin{equation}
u=\frac{\sum_{\lambda}U({\lambda})}{\sum_{\lambda}I({\lambda})},
\label{eq:u}
\end{equation}
\begin{equation}
LP=\frac{\sqrt{\left(\sum_{\lambda}Q(\lambda)\right)^2+\left(\sum_{\lambda}U(\lambda)\right)^2}}{\sum_{\lambda}I({\lambda})},
\label{eq:LP}
\end{equation}
and
\begin{equation}
\chi=\frac{1}{2}\arctan\Biggl(\frac{\sum_{\lambda}U(\lambda)}{\sum_{\lambda}Q(\lambda)}\Biggr).
\label{eq:chi}
\end{equation}
The angle of polarization $\chi$ indicates the direction of the plane of linear polarization with respect to the plane of reference (i.e., the parallel to the nearest solar limb, which is the direction for positive Stokes $Q$)
with positive angles counterclockwise.
We note that
$0^{\circ} \leq \chi < 45^{\circ}$,  $135^{\circ} < \chi <180^{\circ}$ for $q > 0$
and $45^{\circ}\leq\chi\leq135^{\circ}$ for $q \leq 0$
\citep[e.g., Equations (1.8) of ][]{Landi2004}.
The wavelength summation in these equations is over those 
ranges indicated with colors in Figure~\ref{fig:iquv_spectra}. 
The exact pixels included in the integration depend on the  
target and slit position, as discussed in Appendix~\ref{appendix_sumpixel}.
The spatial variations of these quantities for the five spectral ranges 
are shown in Appendix~\ref{appendix_spatial_var}.

We determine the error in these quantities from the error propagation based on the photon noise.
The $q$, $u$, and $LP$ errors are quantified with
\begin{equation}
\sigma_{l}=\frac{1}{a\sqrt{N}},
\label{eq:error}
\end{equation}
where $N$ is the total number of photons and $a=(1-\cos234^\circ)/\pi$ the modulation coefficient \citep{Ishikawa2014SoPh,2022SoPh..297..135S}.
The noise can introduce a non-zero contribution in $LP$, as can be seen from Equation (\ref{eq:LP}).
However, such a contribution can be neglected in this analysis because
the $\sigma_l$ are found to be much smaller than the corresponding amplitudes of $q$, $u$, and $LP$ (see the mean $\sigma_l$ in Table \ref{tab:mgk_wlpixel} and the shaded areas in Figures \ref{fig:kcore_all} - \ref{fig:farr_all}).

The error for $\chi$ is
\begin{equation}
\sigma_{\chi}=\frac{1}{2}\frac{\sigma_l}{LP}.
\end{equation}
The large uncertainty found for certain values of $\chi$ (for example, around $680\arcsec$ in Figure \ref{fig:kcore_all}) is due to the close-to-zero values of $LP$, as can be seen from the equation above.

\begin{figure*}
\includegraphics{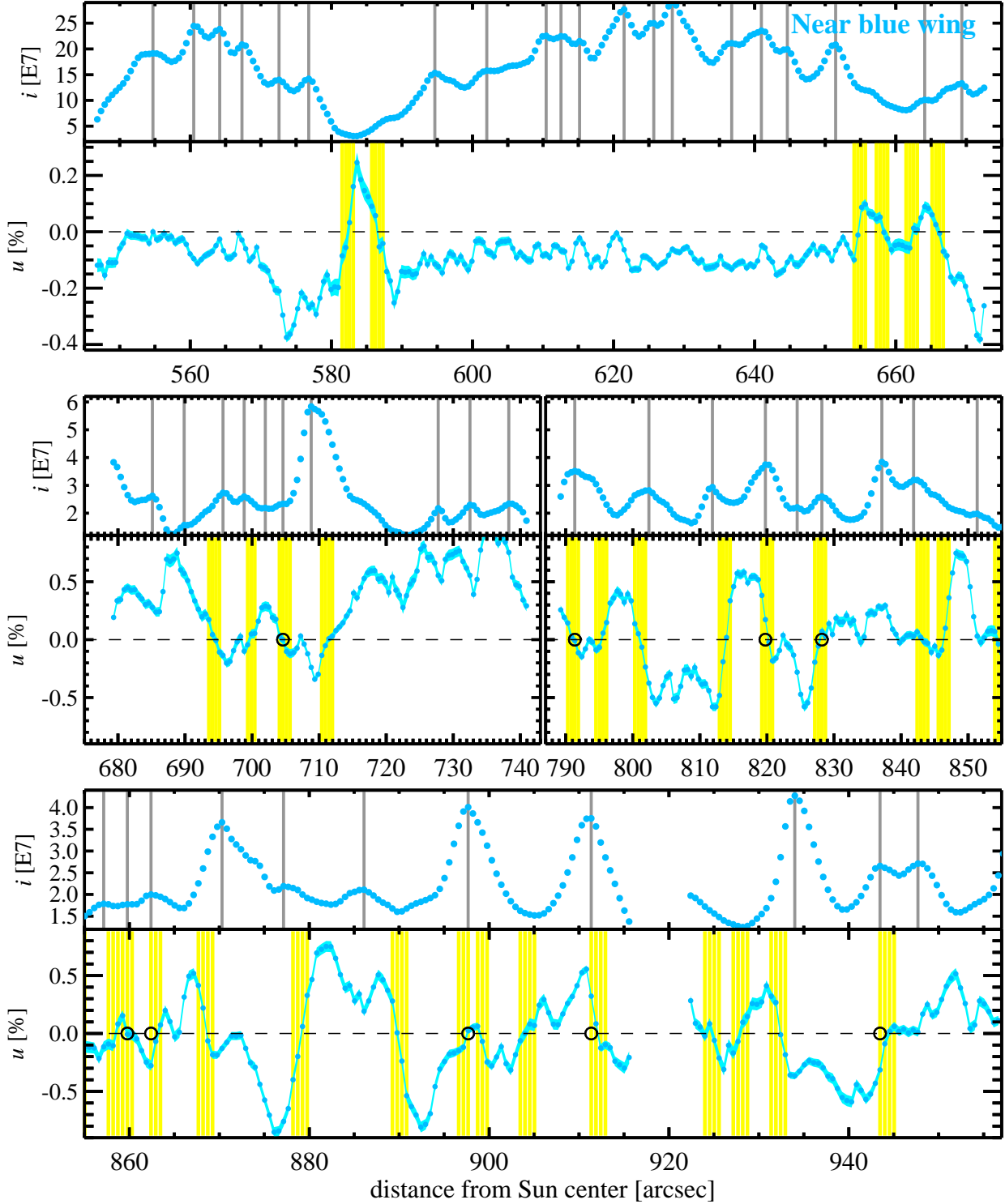}
\caption{Spatial variation of $i$ and $u$ at the near blue wing of the Mg~{\sc ii} $k$ 
line (see the light blue spectral range in Fig.~\ref{fig:iquv_spectra}), 
calculated with Equations (\ref{eq:i}) and (\ref{eq:u}), respectively. 
The shaded area in cyan in the panels for $u$ show $\pm\sigma_l$ errors.
The top panel corresponds to the 
bright region of the plage, while the middle and bottom panels correspond
to the quiet region.
The locations where the intensity presents a local maximum 
are indicated by gray vertical lines,
while the ones where $u$ changes the sign are shown by yellow vertical stripes.
The black circles show the positions 
where these local maxima 
are co-located with
$u$ zero-crossing locations.
\label{fig:i_u_nbw}
}
\end{figure*}

\begin{figure*}
\includegraphics{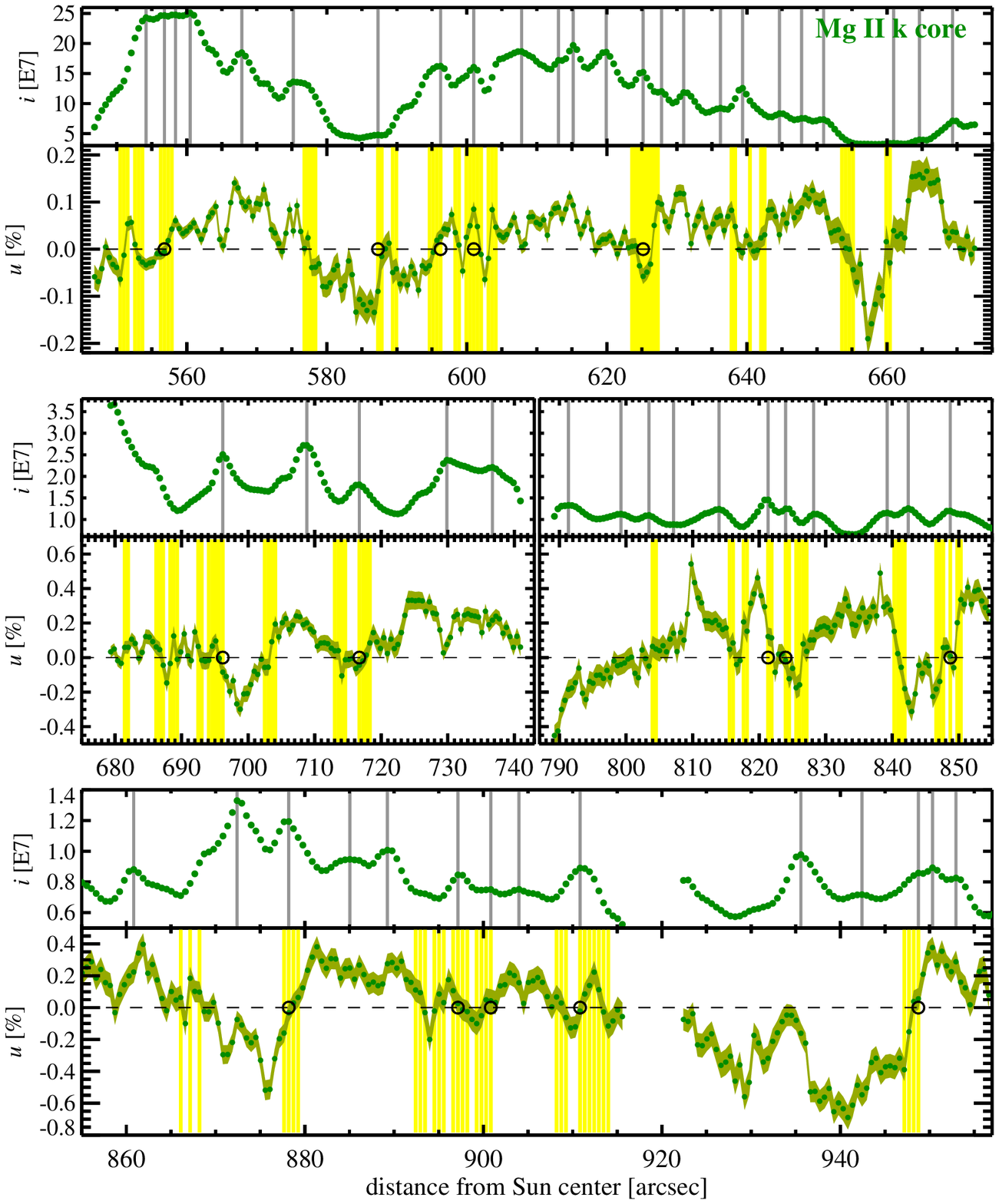}
\caption{Same as Fig.~\ref{fig:i_u_nbw}, but for the core of the Mg~{\sc ii} $k$ line (see green spectral range in Fig.~\ref{fig:iquv_spectra}).
\label{fig:i_u_core}
}
\end{figure*}

\subsection{Longitudinal magnetic field inference}
\label{sec:BL}
We compare the quantities defined above
with the longitudinal 
component of the magnetic field ($B_{L}$) inferred 
at several heights in the solar atmosphere.
We derive $B_{L}$ 
in the plage target
from the Zeeman-induced $V/I$ signals  
observed in the Mg~{\sc ii} $h$ \& $k$ lines 
and in the two Mn~{\sc i} lines
(gray areas in Figure~\ref{fig:iquv_spectra}). 
The $V/I$ profiles of the Mg~{\sc ii} $h$ (effective Land\'e factor $g_\mathrm{eff}=1.33$) and $k$ ($g_\mathrm{eff}=1.167$) lines  
consist of two external lobes and two inner lobes of smaller amplitude 
(see also the bottom panel of Figure~\ref{fig:rgb}).
These external and inner lobes of $V/I$ encode information on the magnetic field 
at the middle and top layers of the chromosphere, respectively \citep{2020ApJ...891...91D}.
The $V/I$ profiles of the two Mn~{\sc i} lines at 279.91~nm ($g_\mathrm{eff}=1.94$) and 280.19~nm ($g_\mathrm{eff}=1.7$) have only two lobes, which
provide information at the bottom of the chromosphere \citep{2022ApJ...940...78D}.

We infer $B_{L}$  by applying the WFA exactly as we did in \cite{2021SciA....7.8406I}.
PRD has an impact on the amplitudes of the external lobes of the $h$ \& $k$ lines and thus the WFA tends to underestimate $B_{L}$ \citep{2016ApJ...831L..15A,2016ApJ...830L..24D}.
Such an underestimation is more significant in the $k$ line.
Moreover, the blue wing of the $k$ line is blended with the Mn~{\sc i} line. 
Therefore, we apply the WFA only to the external lobes of the $h$ line to infer the $B_{L}$ at the middle chromosphere.
The resulting spatial variations of $B_{L}$ along the slit of the spectrograph 
are shown in Appendix~\ref{appendix_spatial_var} (bottom panels of Figures~\ref{fig:kcore_all}$-$\ref{fig:farr_all}).
The $1\sigma$ error of $B_{L}$ ($\sigma_{B}$) is also obtained based on the
uncertainty of $V/I$ due to the photon noise.
Note that no significant $V/I$ signals are detected at the quiet limb target (upper panel of Figure~\ref{fig:iquv_spectra}) and 
thus we only infer $B_{L}$ at the plage target.

\section{Results}
\subsection{Sign change of $u$}
The geometrical complexity of the upper solar chromosphere has a crucial impact on
the scattering polarization of strong resonance lines \citep{2018ApJ...866L..15T}. This 
impact is due to the lack of axial symmetry of the incident radiation field (i.e., the 
radiation that illuminates each point within the medium), caused by the   
horizontal inhomogeneities of the solar atmospheric plasma and/or 
by the gradients of the macroscopic velocities.
In their analyses of the 
CLASP1 observations of the scattering polarization in the hydrogen Lyman-$\alpha$ and in the 
resonance line of Si {\sc iii}, \cite{2017ApJ...841...31I} 
searched for signatures indicating a lack of such axial symmetry. 
They noted that in the wings of the Lyman-$\alpha$ line, where 
the Hanle effect does not operate, the $U/I$ signal tends to change its sign around the spatial locations 
where the Lyman-$\alpha$ intensity at the same 
wavelength shows a local maximum value.       
As schematically shown in their Figure~7 \citep[see also][]{2012ASPC..456...59S}, 
the correlation between the sign change in the $U/I$ wing signal and the local maxima in intensity is consistent with the scattering signals due to the lack of axial symmetry.
Here, we investigate 
whether a similar behavior is present 
across the Mg~{\sc ii} $h$ \& $k$ lines.

Figures~\ref{fig:i_u_nbw} and \ref{fig:i_u_core} show the spatial variation of $i$ (Equation~(\ref{eq:i})) and $u$ (Equation~(\ref{eq:u})) at the near blue wing and the core of the $k$ line (see also Figures~\ref{fig:i_u_fbw}, \ref{fig:i_u_frw}, and \ref{fig:i_u_nrw} in Appendix~\ref{appendix_spatial_iu} for the far blue wing, the far red wing, and the near red wing of the $k$ line, respectively).
We mark the pixels $x_{\mathrm{max}}$ where the $i$ signals 
show a local maximum (gray vertical line) and then
examine whether the $u$ signal changes its sign (i.e., crosses zero) around such spatial locations.
If the signs of $u$ at two neighboring pixels around $x_{\mathrm{max}}$ are opposite
(i.e., $u_{x_{\mathrm{max}}-1}\cdot u_{x_{\mathrm{max}}+1}<0$
or
$u_{x_{\mathrm{max}}-2}\cdot u_{x_{\mathrm{max}}+2}<0$)
we classify this pixel as a $u$ zero-crossing point (black circle in the figure).
However, if the $u$ signal at any of such neighboring pixels
turns out to be smaller than the noise level (e.g., $|u_{x_{\mathrm{max}}-1}|<\sigma_{x_{\mathrm{max}}-1}$) and its sign is opposite to that corresponding to 
the outer neighboring pixel
(e.g., $u_{x_{\mathrm{max}}-1}\cdot u_{x_{\mathrm{max}}-2}<0$),
we do not count it as a zero-crossing point.

Table~\ref{tab:iu_peak} summarizes the number of identified $i$ local maxima 
and $u$ zero-crossing points.
The region at a distance from the Sun center of less than $673\arcsec$
(i.e., the area below the dust in the bottom panel of
Figure~\ref{fig:iquv_spectra}) is what we call 
plage region (top panels in Figures~\ref{fig:i_u_nbw} and \ref{fig:i_u_core}, and in 
Figures \ref{fig:i_u_fbw}, \ref{fig:i_u_frw}, and \ref{fig:i_u_nrw} in Appendix \ref{appendix_spatial_iu}), since at most of the slit positions we find a sizable longitudinal component of the magnetic field.
The remaining regions (i.e., the area above the dust in the  
bottom panel of Figure~\ref{fig:iquv_spectra}
and  the region in the top panel of Figure~\ref{fig:iquv_spectra})
are what we call  
quiet region.
In the quiet region, we find that 
at the considered wavelength ranges
the $u$ signal changes sign in  
more than roughly 30\% of the locations where $i$ shows a local maximum.
In the plage region, some of the locations with 
local maxima are cospatial with a change in the sign of $u$. 
However, 
the fraction of $u$ zero-crossing points associated with local maxima in $i$ is lower than in the quiet Sun region,
except in the case of the Mg~{\sc ii}~$k$ far red wing (see also Figure~\ref{fig:i_u_frw}).

We investigate whether the number of local $i$ maxima which are co-located with a change in the sign of $u$ (hereafter, the number) can be explained by random chance (see Appendix \ref{appendix_u_zero_cross}).
In the quiet region, the number is larger than the mean that would be expected in the random chance scenario by at least $1\sigma$ at all considered wavelength ranges and at the $2\sigma$ level in the core and red wing ranges.
On the other hand, in the plage region, for the $k$ core, near red wing, and far blue wing, the spatial coexistence of the local intensity maxima and $u$ zero-crossing is consistent with random chance.
In the near blue wing, the number is smaller than the mean expected due to random chance by $1\sigma$. 
As in the quiet region, at the far red wing,
the number exceeds the mean in the random chance case by $2\sigma$.

\begin{table*}
\begin{center}
\begin{tabular}{ccccc}
&  \multicolumn{2}{c}{Plage}
&  \multicolumn{2}{c}{Quiet Sun}\\ 
&  \#$i$ max. & \#$u$ zero-cross
&  \#$i$ max. & \#$u$ zero-cross\\
\hline\hline
Mg~{\sc ii} $k$ core (green)      
    & 24 &  5 (21\%) & 30 & 10 (33\%)\\
Mg~{\sc ii} $k$ near blue wing (light blue) 
    & 20 &  0 (0\%) & 31 &  9 (29\%)\\
Mg~{\sc ii} $k$ near red wing (red) 
    & 26 &  4 (15\%) & 30 & 10 (33\%)\\
Mg~{\sc ii} $k$ far blue wing (violet) 
    & 31 & 9 (29\%) & 62 & 23 (37\%) \\
Mg~{\sc ii} $k$ far red wing (magenta) 
    & 33 & 21  (64\%) & 59 & 34 (58\%)\\
\hline
\end{tabular}
\caption{Number of $i$ maxima and 
$u$ zero-crossing
points
for the plage and quiet regions  
at the selected spectral ranges, namely at the core (green in Fig.~\ref{fig:iquv_spectra}),
the near blue and red wings (light blue and red in Fig.~\ref{fig:iquv_spectra}, respectively) and 
at the far blue and red wings (violet and magenta in Fig.~\ref{fig:iquv_spectra}, respectively) 
of the Mg~{\sc ii} $k$ line.
The plage region pixels are those 
at distances from the disk center of $543\arcsec - 673\arcsec$
while the quiet region pixels are those with $679\arcsec - 984\arcsec$.
The values within parenthesis
give the fraction of $u$ zero-crossings with respect to the number of $i$ maxima.}
\label{tab:iu_peak}
\end{center}
\end{table*}

\begin{figure}[ht!]
\plotone{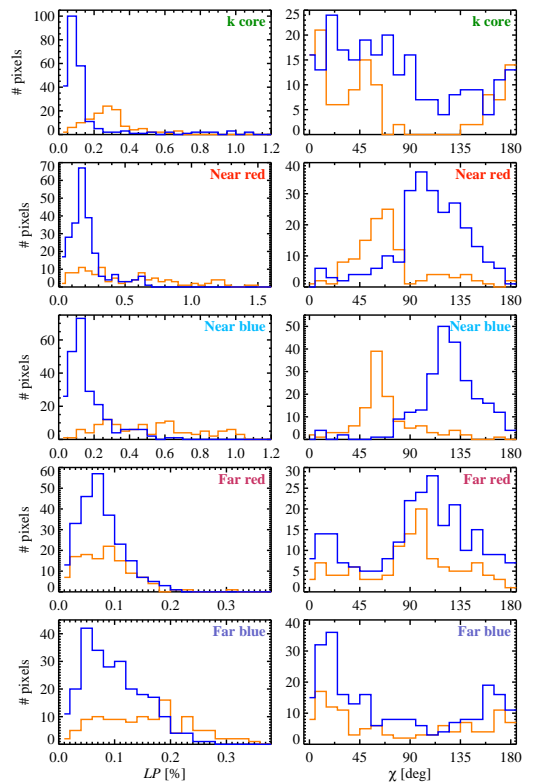}
\caption{Histograms of $LP$ (left column) and $\chi$ (right column) for the quiet (in orange, at the distance from the Sun center of $679\arcsec - 741\arcsec$) and the plage (in blue, at $547\arcsec - 673\arcsec$) regions and the following wavelength ranges of the Mg~{\sc ii}~$k$ line
(from top to bottom): core, near red and blue wings, and far red and blue wings.
}
\label{fig:chi_LP}
\end{figure}

\begin{figure}[ht!]
\plotone{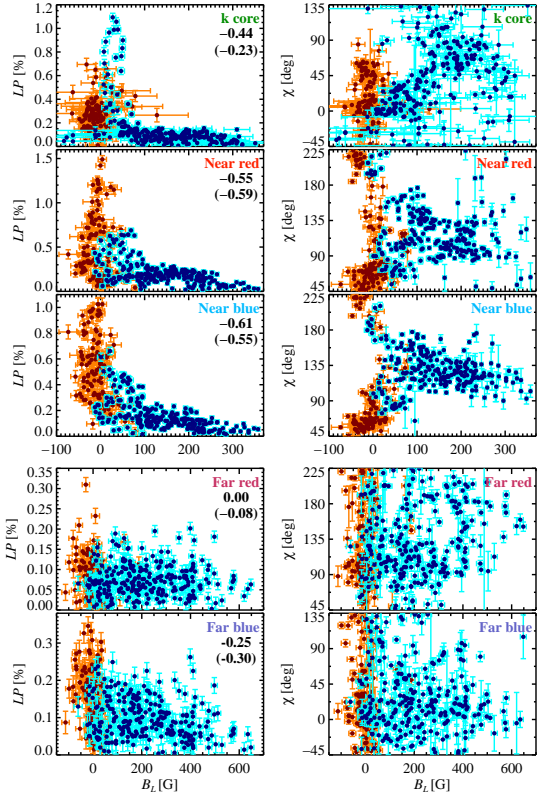}
\caption{Scatter plots of $LP$ vs. $B_{L}$ (left column) and of $\chi$ vs. 
$B_{L}$ (right column) for each of the selected wavelength ranges.
The dark blue points with cyan error bars ($\pm\sigma_{B}$, $\pm\sigma_{l}$, or $\pm\sigma_{\chi}$)
correspond to the plage region at the distance from the Sun center of $547\arcsec - 673\arcsec$, 
whereas the brown points with orange error bars
correspond to 
the quiet region at $679\arcsec - 741\arcsec$.
The numbers shown within
each panel in left column 
are the correlation coefficients between $LP$ and $B_{L}$ 
for the plage region (dark blue points).
The values in parenthesis are the coefficients for the plage pixels with $B_{L}>100$ G.
\label{fig:B_LP}
}
\end{figure}

\subsection{Dependence of $LP$ and $\chi$ on $B_{L}$}
\label{histogram}
In this section, we study how the degree of linear polarization $LP$ (Equation~(\ref{eq:LP})) and the linear polarization angle $\chi$ (Equation~(\ref{eq:chi})) vary with 
$B_{L}$, with the aim of investigating the impact of magnetic fields on the scattering polarization across the Mg~{\sc ii}~$h$ and $k$ lines.
In the quiet Sun target close to the limb,
the lack of $V/I$ signals does not allow for the inference of the longitudinal component of the magnetic field.
Therefore, we limit our investigation to the plage 
target pixels and classify them into 
two groups depending on the distance from the Sun center: $574\arcsec-673\arcsec$ as the plage 
region (below the dust in the lower panel 
of Fig.~\ref{fig:iquv_spectra}) and 
$679\arcsec - 741\arcsec$ 
as the quiet region (above the dust in the lower panel of Fig.~\ref{fig:iquv_spectra}).

\subsubsection{$LP$ distribution}
The left column of Figure~\ref{fig:chi_LP} shows the $LP$ histograms of the plage (blue) 
and the quiet (orange) regions for the previously indicated 
five wavelength ranges.
In most of the wavelength ranges, the quiet region shows a 
wider distribution of $LP$ values (for example, from 0\% to 1.5\% at the near red wing) than in the plage region. 
Moreover, the quiet region histograms
for the near red and blue wings and for the far blue wing
do not show any significant peak.
In contrast, in the plage region, each of the $LP$ histograms shows a clear peak at 
relatively low $LP$ values. Especially at the core, near red and blue 
wings of the Mg~{\sc ii} $k$ line, 
the $LP$ distributions
are clearly different between the quiet and plage regions,
clearly indicating that the weakly magnetized region tends to have larger $LP$ signals.

The left column of Figure~\ref{fig:B_LP}, which shows the scatter plots 
of $LP$ vs. $B_{L}$ (Section~\ref{sec:BL}) for the quiet (brown points with orange error bars) 
and the plage (dark blue with cyan error bars) regions, for the five selected spectral ranges, 
allows to investigate the behavior of $LP$ with $B_{L}$.
We compare $LP$ with $B_{L}$ at the approximate height of formation of each spectral range indicated with a color in Figure~\ref{fig:iquv_spectra}, 
namely, $LP$ at the core of the Mg~{\sc ii}~$k$ line vs. $B_{L}$ at the top of the chromosphere (estimated via the inner lobes of $V/I$ of the Mg~{\sc ii}~$h$ \& $k$), 
$LP$ at the near blue and red wings of the Mg~{\sc ii}~$k$ vs. $B_{L}$ at the middle chromosphere (estimated via the external lobes of $V/I$ of the Mg~{\sc ii}~$k$), 
and $LP$ at the far blue and red wings vs. $B_{\mathrm{L}}$ at the bottom chromosphere (estimated via the $V/I$ signals of the Mn~{\sc i} lines).
It can be observed that
at the Mg~{\sc ii}~$k$ core, the pixels 
with the largest  $LP$ belong to the plage region
although their field strength is relatively weak ($B_{L}<50$~G).
The high $LP$ region extends for $10\arcsec$, showing its peak at 657\arcsec (see fourth panel from the top in Figure \ref{fig:kcore_all}). Such an enhancement of $LP$ comes from the strong $Q/I$ signal and is visible only in the $k$ core (see $Q/I$ spectra in Figure \ref{fig:iquv_spectra}).
In this region, the Mg {\sc ii} intensity spectra show clear self-absorption at the core.
We attribute the large-amplitude $Q/I$ signals found at such spatial points to scattering polarization and we note that, if they originated from the Zeeman effect due to the transverse component of the magnetic field, similar linear polarization signals should also be detected in the $h$ line. 

There seems to be a negative correlation 
between $LP$ and $B_{L}$ (i.e., $LP$ is the smaller for stronger $B_{L}$) 
at the core and the near red and blue wings of $k$ line
as shown by the linear Pearson correlation coefficients given inside each panel.
In the near blue and red wings, the coefficients remain similar  
when considering only the pixels where $B_{L}>100$ G, but it becomes smaller in the core (see the values in the parenthesis).
On the other hand, at the far red and blue wings of the Mg~{\sc ii}~$k$ line 
(bottom two panels of Figure~\ref{fig:B_LP}), $LP$ varies over a wide range of values 
regardless of the $B_{L}$ value, with no significant differences 
between the quiet and plage regions: there is no evidence of correlation between $LP$ and $B_{L}$ in neither the plage or quiet regions. Indeed, the correlation coefficients 
between $LP$ and $B_{L}$ are $0.0$ at the far red wing and $-0.25$ 
at the far blue wing, which are smaller than at the core and near wings of the $k$ line
and do not change very much for the pixels with $B_{L}>100$ G.

\subsubsection{$\chi$ distribution}
The right column of Figure~\ref{fig:chi_LP} shows 
the $\chi$ histograms for the five selected spectral ranges.
Note that there is a $180^{\circ}$ ambiguity in $\chi$ ($\chi = \chi + 180^{\circ}\times m$ with $m=0,\pm1,\pm2, \ldots$).
Radiative transfer calculations 
in an unmagnetized semi-empirical 1D model atmosphere predict $U=0$ and $Q>0$ (i.e., $\chi=0^{\circ}$) at the core, $Q<0$  (i.e., $\chi=90^{\circ}$) at the near blue and red wings, $Q>0$ (i.e., $\chi=0^{\circ}$) at the far blue wing, and $Q<0$ (i.e., $\chi=90^{\circ}$) at the far red wing of the $k$ line \citep{Belluzzi2012ApJ,2016ApJ...830L..24D,2016ApJ...831L..15A}.
The quiet Sun histograms shown in orange are roughly consistent with this prediction.
For example, the histogram of the $k$ core shows high frequency around $0^{\circ}$ or  $180^{\circ}$.
The peaks of the quiet Sun histograms for the near blue and red wings deviate from $90^{\circ}$ by $20^{\circ}-30^{\circ}$. 

At the far red and blue wings (bottom two panels), the $\chi$ histograms in the plage region
are similar to those in the quiet region.
In the plage region there also seems 
to be a preference in $\chi$ for $\sim90^{\circ}$ at the far red wing and for $\sim0^{\circ}$ at the far blue wing, respectively, as in the quiet region.
However, at the near red and blue wings and the core (upper three panels), 
the quiet and plage region histograms are different. 
At the near red and blue wings,
most of the quiet region pixels have $0^\circ\le\chi\le90^\circ$, while most of the plage region pixels have $90^\circ\le\chi\le180^\circ$.
At the core of the $k$ line, the plage region pixels show a more uniform distribution, with a 
non-negligible number of pixels with $67.5^\circ\le\chi\le135^\circ$, while this behavior 
is not found in the quiet Sun pixels.

The right column of Figure~\ref{fig:B_LP} shows the scatter plots of $\chi$ vs. $B_{L}$.
The assignment of $B_{L}$ for the comparison with each selected spectral range is the same as in the left column of  Figure~\ref{fig:B_LP}.
The linear polarization angle $\chi$ has a $180^{\circ}$ ambiguity 
and we choose the plot range so that the majority of the plage 
pixels can smoothly cluster in each panel (i.e., $-45^\circ\leq\chi\leq135^\circ$ 
for the core and the far blue wing of the $k$ line, and $45^\circ\leq\chi\leq225^\circ$ 
for the near red, near blue, and far red wings).  
In the core of the $k$ line, the quiet region distribution (brown points) lies 
around $0^\circ$, although the variation is relative large (i.e., roughly $\pm45^\circ$).
The plage region pixels, 
where $|B_{L}|<100$~G,
are also clustered around $0^{\circ}$. However, 
in the pixels where $B_{L}$ is larger,
they scatter over a wide range of $\chi$ with some preference for $45^{\circ}<\chi<90^{\circ}$ at $150<B_{L}<250$ G. 

At the near blue and red wings,
most of the quiet region pixels are clustered at 
$45^{\circ}\le\chi\le90^{\circ}$ and $180^{\circ}\le\chi\le225^{\circ}$, which correspond to  $0^{\circ}\le\chi\le90^{\circ}$ in Figure \ref{fig:chi_LP}.
However, the behavior of $\chi$ in the plage is very different.
The plage pixels with 
$B_{L}\lesssim50$ G show $\sim90^{\circ}$ and $\sim{180}^{\circ}$, 
which are at the outer edges of the clusters of the quiet region pixels ($45^{\circ}\le\chi\le90^{\circ}$ and $180^{\circ}\le\chi\le225^{\circ}$). 
Moreover, most of the plage pixels with stronger field of $B_{L}>100$ G tend to concentrate at $90^{\circ}\le\chi\le135^{\circ}$. Finally, at the 
far blue and red wings, we cannot find any clear characteristic behavior with $B_{L}$ in either the quiet or plage regions.

In summary, at the core and the near wings of the Mg~{\sc ii} $k$ line, 
the $LP$ and $\chi$ distributions are significantly different  
between quiet and plage regions, and these properties seem to be dependent on $B_{L}$.
On the other hand, at the far wings of the Mg~{\sc ii} $k$ line, 
we do not find any difference nor dependence on $B_{L}$ as clear as 
at the core and the near wings.

\section{Discussion}
CLASP2 succeeded in obtaining 
the first spatially and spectrally resolved
measurements of the 
linear and circular polarization signals across the solar 
Mg~{\sc ii} $h$ \& $k$ lines. 
The Zeeman-induced circular polarization signals have allowed us  
to determine the longitudinal component ($B_{L}$) of the magnetic field 
at three different heights of the solar chromosphere 
(top, middle, and bottom\footnote{The bottom chromosphere is the region just above the temperature minimum in standard semi-empirical models. The top chromosphere is just below the height at which the temperature begins to rise dramatically. The middle chromosphere is the intermediate zone between them.}) via the application of the WFA 
to the Mg~{\sc ii} $h$ \& $k$ and the Mn~{\sc i} lines \citep{2021SciA....7.8406I}. 
Recently, by comparing the CLASP2 observations 
with the theoretical calculations of \cite{2012ApJ...750L..11B}, it has been 
confirmed that PRD and $J$-state interference play a crucial role 
in producing the scattering polarization signals across the Mg~{\sc ii} $h$ \& $k$
lines \citep{2022ApJ...936...67R}. 
The linear polarization signals of these resonance lines 
encode information on the magnetic fields of the solar chromosphere through 
the Hanle and MO effects.
Thanks to the unique data provided by CLASP2, in this paper we have investigated  
the linear polarization signals observed 
at different spectral ranges across the Mg~{\sc ii} $h$ \& $k$ lines, 
comparing them with our estimation of 
$B_{L}$ at their approximate heights of formation in the solar 
chromosphere. 

Our aim has been to clarify the impact 
of the magnetic field on the linear polarization signals across these ultraviolet lines.
Depending on the selected wavelength range, 
the effects can be different, 
because of the different heights of formation and the different mechanisms that 
introduce magnetic sensitivity in the core and wings of the lines.
Thus, we have analyzed separately the linear polarization signals at the $k$-line core, 
at the near blue and red wings, and at the far blue and red wings of the Mg~{\sc ii}~$k$ line. 
We point out that an analysis of the amplitude of the $Q/I$ profile alone is not sufficient to judge on the possible operation of the Hanle and the MO effects.
Such amplitude heavily depends on the anisotropy of the radiation field, which in turn primarily depends on the temperature stratification of the atmosphere in the observed region. The latter may strongly vary moving from a quiet region to a plage. 
Indeed, while the emergent intensity in the 1D semi-empirical model P of \cite{Fontenla1993}, which is representative of a plage region, is significantly larger than in their quiet Sun model C due to its enhanced temperature below the transition region, the P model shows less fractional linear polarization \citep{2017SSRv..210..183T,2020ApJ...891...91D} because of the smoother temperature gradient \citep{2018SoPh..293...74I}. Thus, we have investigated also the $U/I$ and the behavior of the linear polarization angle $\chi$.

First, we focus on the spatial variation of the $U/I$ signals, 
which are induced by the presence of the
magnetic fields (via the Hanle and MO effects)  
and by the lack of axial symmetry of the radiation field. 
In the quiet region, where $B_{L}$ is relatively low, we find that in about  
30\% of the bright structures (local intensity maxima)
the $U/I$ signals change their sign (i.e., $U/I$ crosses zero), both at the core of the $k$ line
(where the Hanle effect operates) and at the wings (where the MO effects operate).
The numbers of spatial coincidences between the local intensity maxima and the $U/I$ zero-crossings are larger than those expected from random coincidence, at $1-2\sigma$ levels depending on the considered wavelength range.
However, in the plage region, where $B_{L}$ is substantial, reaching up to 350/650 G at the top/bottom chromosphere,
the fraction of bright structures where 
the $U/I$ signals change their sign 
seem to be smaller,
except in the far red wing of the Mg~{\sc ii}~$k$ line.
Indeed, in the plage region, the number of spatial coincidences is consistent with the number that would be expected from the random chance case (core, near red wing, far blue wing) or it is smaller at $1\sigma$ significance (near blue wing).
In the plage region, the number of spatial coincidences in the far red wing exceeds the expected random chance mean at $2\sigma$ significance.

Next, we have investigated whether the behavior of the  
total linear polarization ($LP$) and the linear polarization angle ($\chi$) 
is different between the quiet and plage regions and whether it varies with $B_{L}$. 
Interestingly, we find that at all the selected wavelength ranges,
$LP$ is significantly suppressed in the plage region. 
Moreover, especially at the core and the near 
wings of the $k$ line, we find anticorrelation between $LP$ and $B_{L}$.
At the core and the near wings of the Mg~{\sc ii} $k$ line, 
the $\chi$ distribution is also different in the quiet and plage regions 
especially in the pixels where $B_{L}>50$ G.
On the other hand, at the far wings of the $k$ line, the $\chi$ distributions in the quiet and plage regions are similar and we do not find any dependence of $\chi$ on $B_{L}$. 

The smaller fraction of $U/I$ zero-crossing points in the plage region 
may be interpreted as due mainly to the influence of the magnetic fields through the Hanle and MO effects, 
dominating over that caused by the lack of axial symmetry.
The smaller $LP$ and the change of $\chi$ 
can be interpreted as 
manifestations of the depolarization and rotation of the plane of linear polarization 
caused by the Hanle and MO effects. 
These three observational signatures are found in the plage region at the core and the near wings of the Mg~{\sc ii} $k$ line, and we consider them as evidence for  
the operation of the Hanle and MO effects.
Note that any change of the linear polarization signals found in this analysis is not due to the transverse Zeeman effect.
If this was the case, we should see corresponding signals in the Mg {\sc ii} $h$ line, which is more sensitive to the Zeeman effect than the $k$ line because it has a larger effective Land\'e factor.

In the $\chi$ vs. $B_{L}$ scatter plots of Figure \ref{fig:B_LP}, the pixels with $B_{L}>100$ G tend to concentrate at $45^{\circ}<\chi<90^{\circ}$ in the core of the $k$-line 
and at $90^{\circ}<\chi<135^{\circ}$ in the near wings of the $k$ line.
For field strengths $B>100$ G, the Hanle effect at the $k$-line core is in its saturation regime and the linear polarization signal is sensitive only to the orientation of the magnetic field \citep[see the two top panels of Figure 9 of][]{2020ApJ...891...91D}.
The MO effects are still sensitive to the magnetic field strength and orientation for such 
relatively strong fields, although the sensitivity to the strength is not as high as for weaker fields. 
Thus, these concentrations around $B_{L}>100$ G in the 
$\chi$ vs. $B_{L}$ 
scatter plots may be compatible with the typical 
strong magnetic field configuration of plage regions.
In the core of the $k$ line, for the plage pixels with $B_{L}>100$ G, the anti-correlation between $LP$ and $B_{L}$ is less significant, and $LP$ seems to be completely suppressed.
This property would be consistent with the Hanle saturation regime.

At the far wings, which originate in the upper photosphere, 
we find only one observational signature suggestive of the operation of the MO effects (i.e., small $LP$ in the plage region)
and evidence for the operation of the MO effects as clear as that found for the near wings could not be obtained.
In general, the impact of MO effects is stronger in the near wings than in the far wings where the relative impact of the continuum is stronger \citep[e.g., Section 5 of][]{2018ApJ...854..150A}. 
Moreover, in the Mg~{\sc ii} $h$ \& $k$ lines, as clearly shown in the response function to the magnetic field in Figure 8 of \cite{2020ApJ...891...91D}, the stronger response occurs higher in the atmosphere (i.e., near wings of the $k$ line) than lower in the atmosphere (i.e., far wings of the $k$ line) via the MO effects.

In this paper, we have provided purely observational evidence on the operation of 
the Hanle and MO effects in the scattering polarization of the Mg {\sc ii} $h$ \& $k$ lines. 
The theoretical modeling of the unprecedented spectropolarimetric data provided by CLASP2 
will be the subject of future papers. To that end, new plasma diagnostic tools are being 
developed, such as a new Stokes inversion code \citep{2022ApJ...933..145L} and a spectral synthesis code that takes into account the effects of horizontal radiative transfer. 
Such ongoing developments are important steps to infer the full vector magnetic field in the plage chromosphere through the Zeeman, Hanle, and MO effects in the near-UV region of the Mg {\sc ii} resonance lines.  

\acknowledgments
CLASP2 is an international partnership between NASA/MSFC, NAOJ, JAXA, IAC, and IAS; additional partners include ASCR, IRSOL, LMSAL, and the University of Oslo.
The Japanese participation was funded by JAXA as a Small Mission-of-Opportunity Program, JSPS KAKENHI Grant numbers JP25220703 and JP16H03963, 2015 ISAS Grant for Promoting International Mission Collaboration, and by 2016 NAOJ Grant for Development Collaboration. The USA participation was funded by NASA Award 16-HTIDS16\_2-0027. 
The Spanish participation was funded by the European  Research Council (ERC) under the European Union's Horizon 2020 research and innovation programme (Advanced Grant agreement No. 742265). 
The French hardware participation was funded by CNES funds CLASP2-13616A and 13617A.
L.B. acknowledges the funding received through SNSF grants 200021-175997 and CRSII5-180238.
The CLASP2 team acknowledges Dr. Shin-nosuke Ishikawa, who led the development 
of the critical component of the polarization modulation unit (PMU) at ISAS/JAXA.
R.I. acknowledges the NAOJ Overseas Visit Program for Young Researchers (FY2019) 
and a Grant-in-Aid for Early-Career Scientists JP19K14771. 

\appendix
\section{wavelength ranges for integration}
\label{appendix_sumpixel}
\begin{figure}
\includegraphics[angle=90,width=\linewidth]{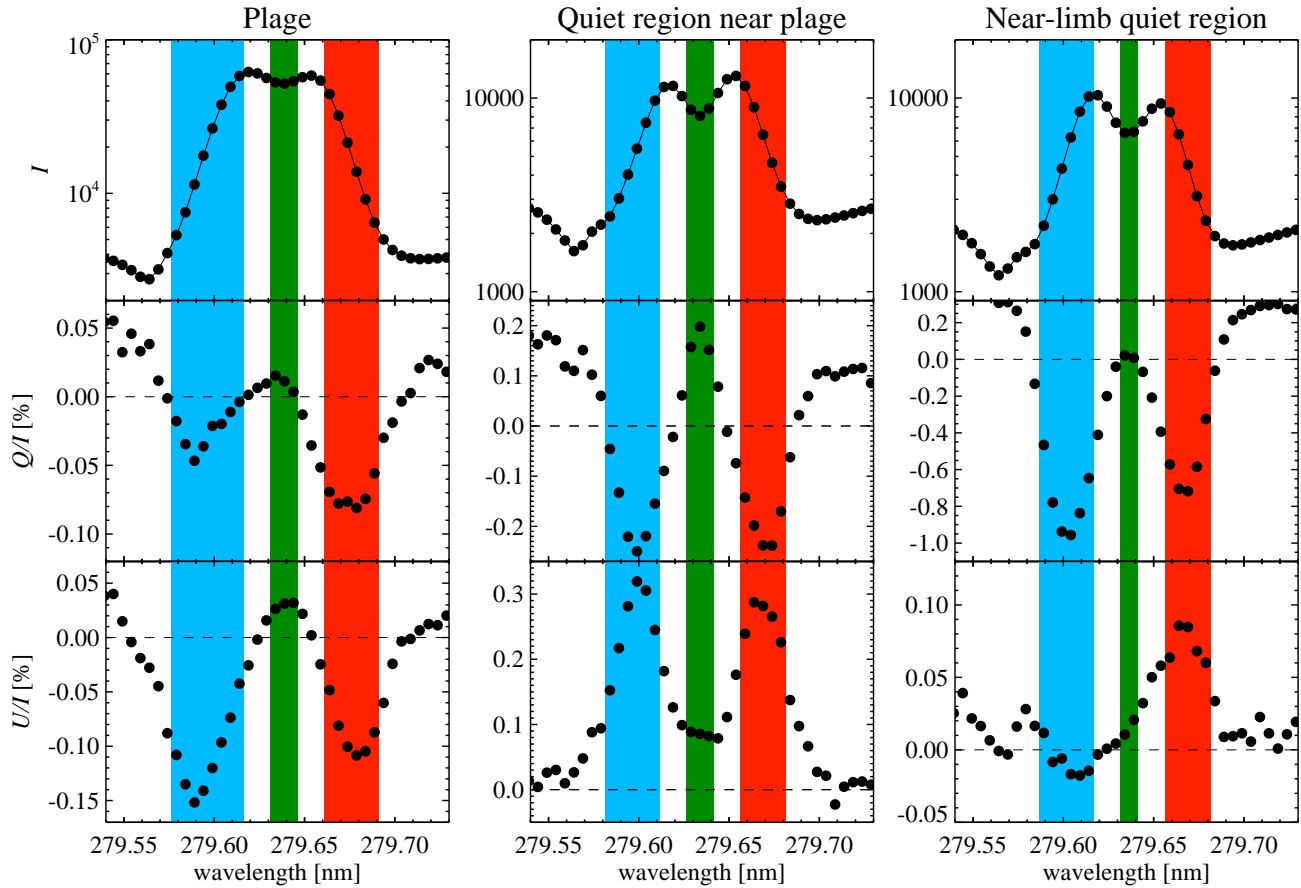}
\caption{From top to bottom: Spatially and temporally averaged Stokes $I$, $Q/I$, and $U/I$ profiles for the plage (left), the quiet region near the plage (middle), and the near-limb quiet region (right).
Photon noise is negligibly small as a result of the spatial averaging over relatively large sections of the slit and thus is not shown here.
The light blue, red, and green colored bands show the wavelength ranges for the near blue and red wings and the core of the $k$ line, respectively, and the same as those indicated in Fig.~\ref{fig:iquv_spectra}.
\label{fig:ave_iqu}
}
\end{figure}

We use temporally and spatially averaged Stokes spectra 
to select the wavelength ranges 
where  the core and near wing signals of the 
Mg~{\sc ii}~$k$ line are integrated (Equations~(\ref{eq:i}$)-($\ref{eq:chi})), and
we distinguish between 
the plage region (at the distance from the Sun center of $547\arcsec - 673\arcsec$, below the dust of the plage target shown in the bottom panel of Figure \ref{fig:iquv_spectra}), the quiet region near the plage ($679\arcsec - 741\arcsec$, above the dust of the plage target), and the quiet region of the limb target ($789\arcsec - 984\arcsec$, upper panel of Figure \ref{fig:iquv_spectra}, except for the dust).
The resulting averaged spectra are shown in Figure~\ref{fig:ave_iqu}.
Note that after deriving the spatially and temporally averaged $I$, $Q$, and $U$, we take the ratio to derive the $Q/I$ and $U/I$ profiles.

Two or three pixels were chosen for studying the signals at the core of the Mg~{\sc ii}~$k$ line (the area highlighted in green).
The light blue and red colored bands refer to 
the near blue and near red wings pixels of the Mg~{\sc ii}~$k$ 
line, which are selected to fully cover the $Q/I$ and $U/I$ lobes in the wings.
The selected wavelength ranges for each region are given in Table~\ref{tab:mgk_wlpixel}.

\begin{table}
\begin{center}
\begin{tabular}{lccc}
\hline \hline
& plage & quiet region near plage & near-limb quiet region\\ 
position & $547\arcsec - 673\arcsec$ & $679\arcsec - 741\arcsec$ & $789\arcsec - 984\arcsec$\\ \hline \hline
Mg~{\sc ii}~$k$ core & $-0.005\le\lambda_{k}\le0.005$~nm & $-0.005\le\lambda_{k}\le0.005$~nm & $-0.0025\le\lambda_{k}\le0.0025$~nm\\
& (3 pixels) & (3 pixels) & (2 pixels)\\
& $\overline{\sigma_{l}}=0.020\%$ &  $\overline{\sigma_{l}}=0.046\%$ &  $\overline{\sigma_{l}}=0.067\%$\\ \hline
Near blue wing & $-0.060\le\lambda_{k}\le-0.025$~nm & $-0.050\le\lambda_{k}\le-0.025$~nm & $-0.047\le\lambda_{k}\le-0.022$~nm\\
& (8 pixels)& (6 pixels)& (6 pixels)\\
& $\overline{\sigma_{l}}=0.017\%$ &  $\overline{\sigma_{l}}=0.042\%$ &  $\overline{\sigma_{l}}=0.042\%$\\ \hline
Near red wing & $0.025\le\lambda_{k}\le0.05$~nm & $0.025\le\lambda_{k}\le0.045$~nm & $0.022\le\lambda_{k}\le0.042$~nm\\ 
& (6 pixels)& (5 pixels)& (5 pixels)\\ 
& $\overline{\sigma_{l}}=0.022\%$ &  $\overline{\sigma_{l}}=0.040\%$ &  $\overline{\sigma_{l}}=0.049\%$\\
\hline
far blue wing & \multicolumn{3}{c}{$279.37 \le \lambda \le279.44$~nm (16 pixels)} \\ 
& $\overline{\sigma_{l}}=0.024\%$ &  $\overline{\sigma_{l}}=0.027\%$ &  $\overline{\sigma_{l}}=0.032\%$\\ \hline
far red wing & \multicolumn{3}{c}{$279.97 \le \lambda \le280.12$~nm (31 pixels)} \\ 
& $\overline{\sigma_{l}}=0.016\%$ &  $\overline{\sigma_{l}}=0.018\%$ &  $\overline{\sigma_{l}}=0.021\%$\\
\hline
\end{tabular}
\caption{Summary of the selected spectral ranges for the core, the near red and blue wings, and the far blue and red wings of the Mg~{\sc ii}~$k$ line. 
The position is the distance from the Sun center 
(see Fig~\ref{fig:iquv_spectra}). $\lambda_{k}$ is
the wavelength with respect to the line center of the averaged Mg~{\sc ii}~$k$ 
intensity profile in each region (i.e., plage, quiet Sun near the plage, and quiet Sun near the limb, see Fig.~\ref{fig:ave_iqu}),  
while $\lambda$ is the absolute wavelength.
The number of selected wavelength pixels is shown in parenthesis for each spectral interval.
The mean value of the uncertainty of $q$, $u$, and $LP$ (Equation (\ref{eq:error})) is derived at each wavelength range and at each position ($\overline{\sigma_{l}}$). Note that $\overline{\sigma_{l}}$ for the near-limb quiet region takes into account the pixels at less than $950\arcsec$, i.e., removing the off-limb pixels.
}
\label{tab:mgk_wlpixel}
\end{center}
\end{table}

\section{Scattering polarization signals and $B_{L}$}
\label{appendix_spatial_var}
\begin{figure*}
\plotone{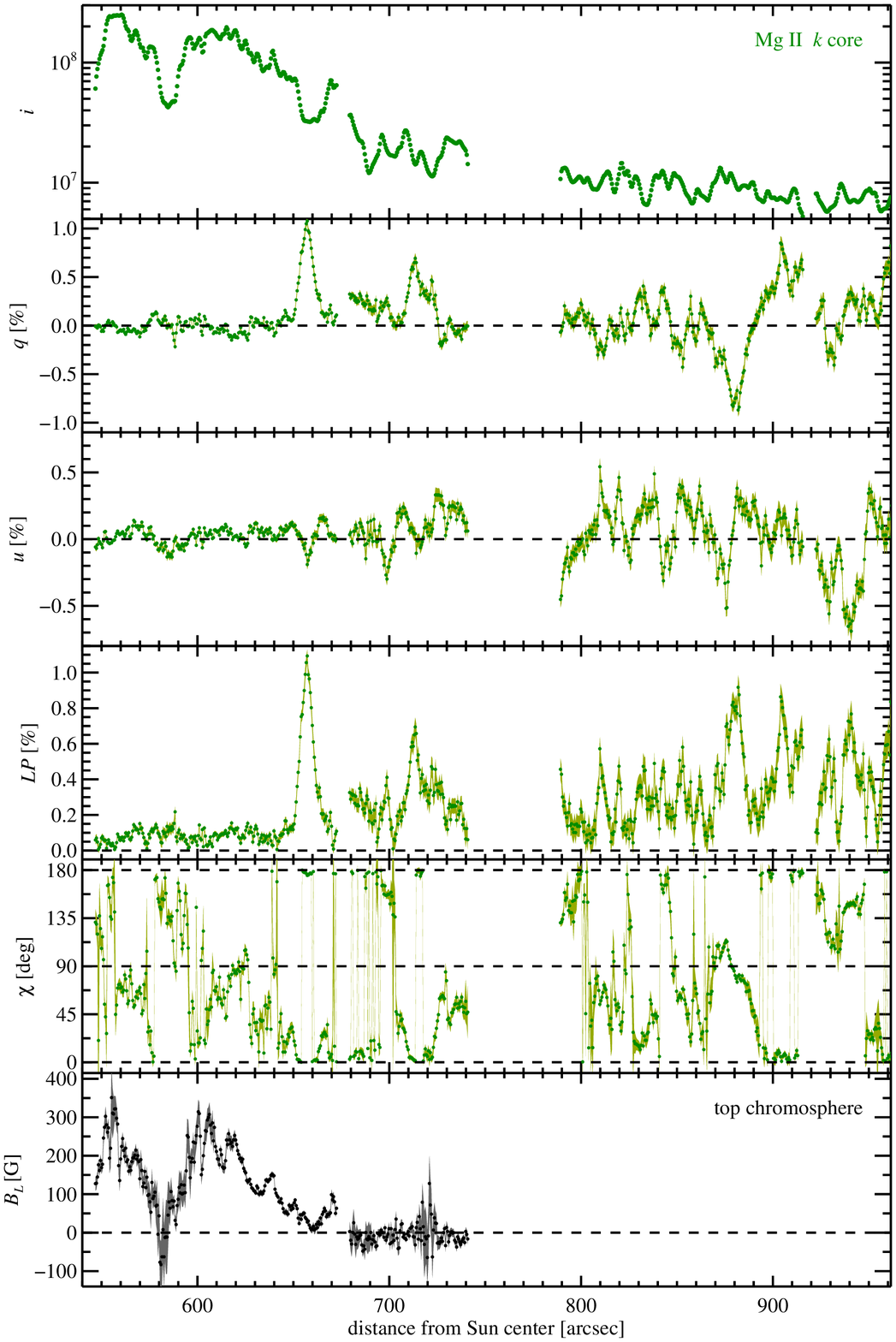}
\caption{From top to bottom: spatial variations of $i$, $q$, $u$, $LP$, and $\chi$ at the core of the Mg~{\sc ii} $k$ line (the wavelength range indicated in green color in Fig.~\ref{fig:iquv_spectra}) and the longitudinal magnetic field $B_\mathrm{L}$ at the top of the chromosphere derived from the inner lobes of the Mg~{\sc ii}~$h$ \& $k$ line. 
The shaded areas show the $\pm1\sigma$ errors ($\pm\sigma_{l}$ for $q$, $u$, and $LP$, $\pm\sigma_{\chi}$ for $\chi$, and $\pm\sigma_{B}$ for $B_{L}$).
\label{fig:kcore_all}
}
\end{figure*}

\begin{figure*}
\plotone{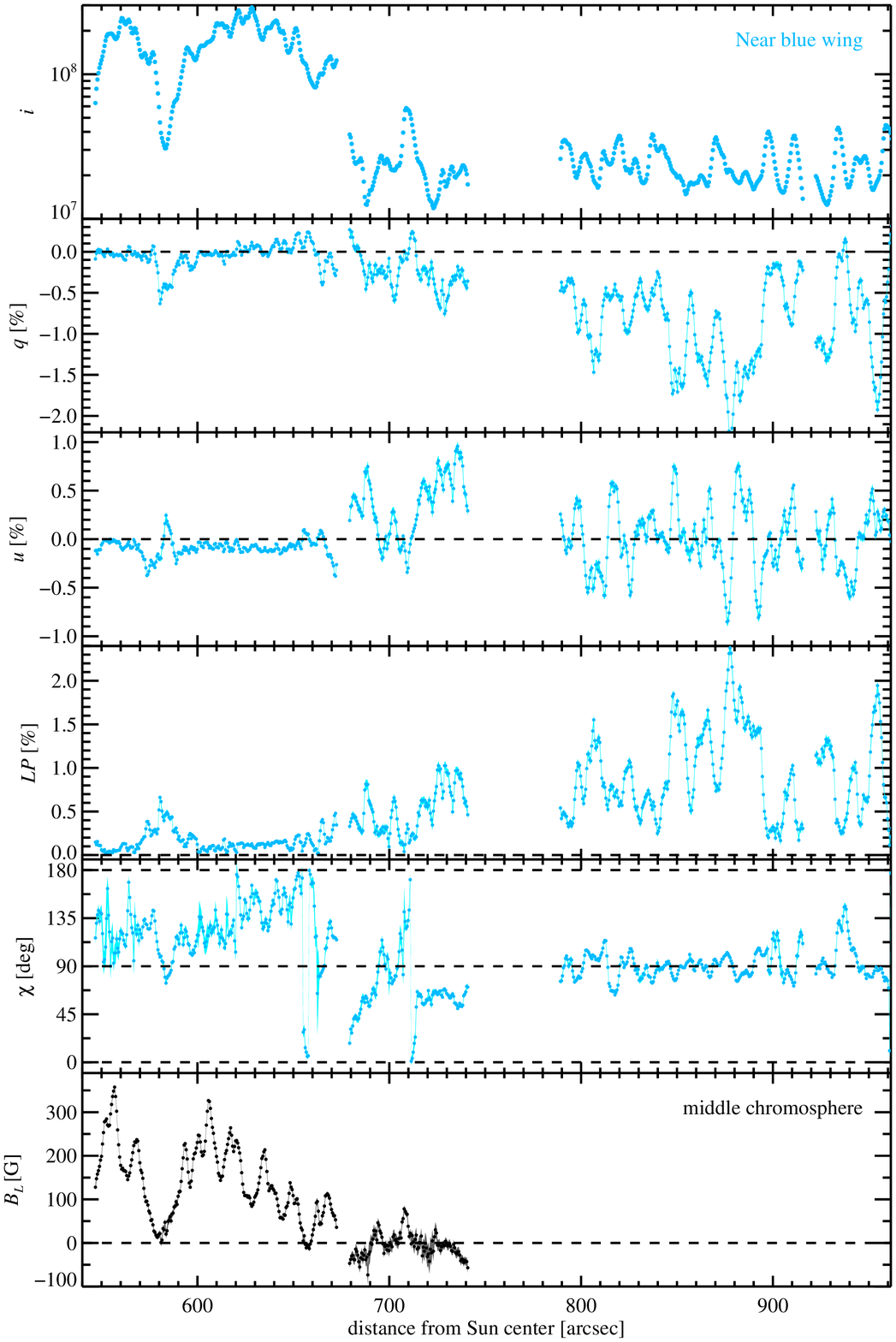}
\caption{From top to bottom: spatial variations of $i$, $q$, $u$, $LP$, and $\chi$ at the near blue wing of the Mg~{\sc ii} $k$ line (the wavelength range indicated in light blue color in Fig.~\ref{fig:iquv_spectra}) and the longitudinal magnetic field $B_\mathrm{L}$ at the middle of the chromosphere derived from the external lobes of the Mg~{\sc ii}~$h$ line.
The shaded areas show the $\pm1\sigma$ errors ($\pm\sigma_{l}$ for $q$, $u$, and $LP$, $\pm\sigma_{\chi}$ for $\chi$, and $\pm\sigma_{B}$ for $B_{L}$).
\label{fig:nearb_all}
}
\end{figure*}

\begin{figure*}
\plotone{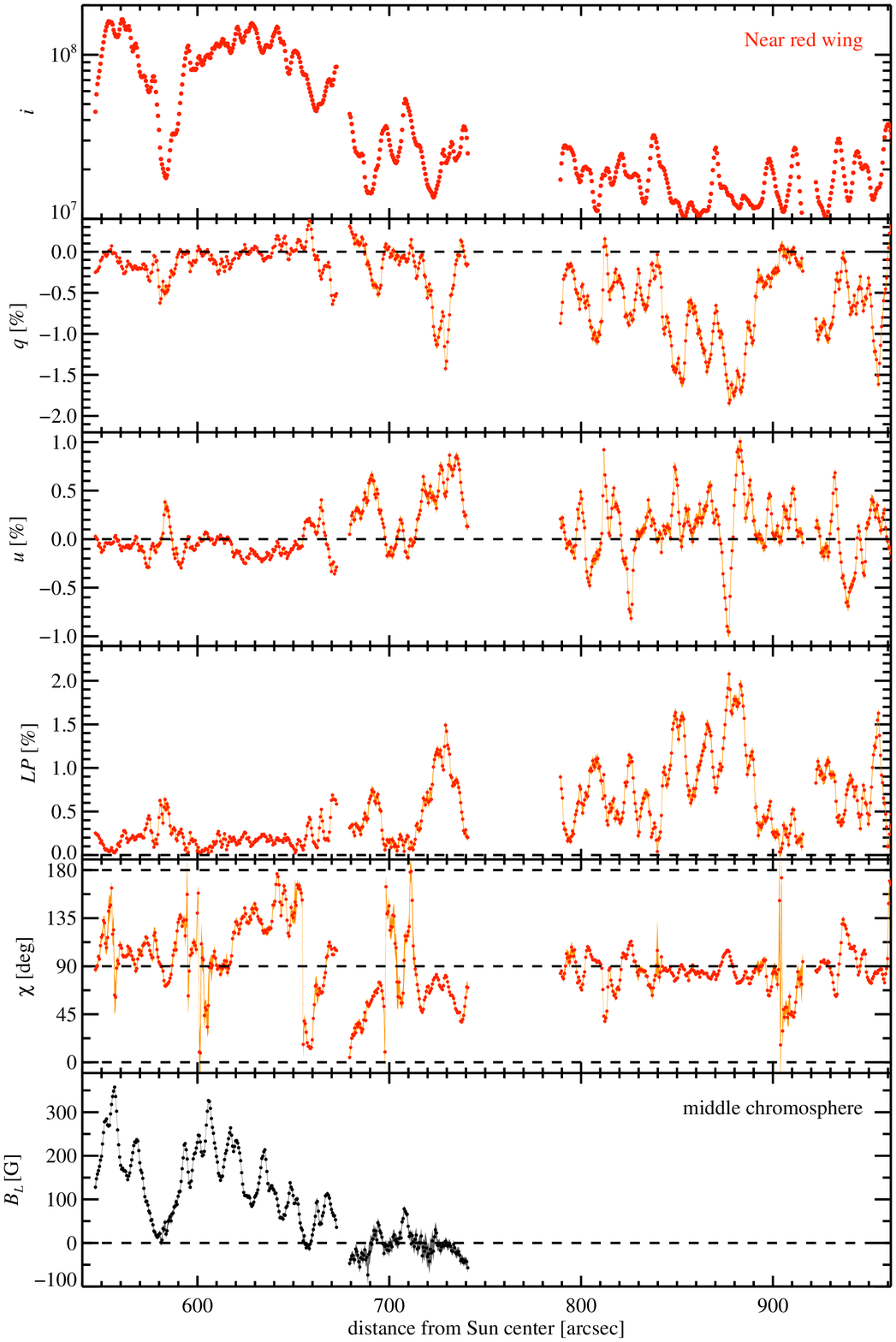}
\caption{From top to bottom: spatial variations of $i$, $q$, $u$, $LP$, and $\chi$ at the near red wing of the Mg~{\sc ii} $k$ line (the wavelength range indicated in red color in Fig.~\ref{fig:iquv_spectra}) and the longitudinal magnetic field $B_\mathrm{L}$ at the middle of the chromosphere derived from the external lobes of the Mg~{\sc ii}~$h$ line.
The shaded areas show the $\pm1\sigma$ errors ($\pm\sigma_{l}$ for $q$, $u$, and $LP$, $\pm\sigma_{\chi}$ for $\chi$, and $\pm\sigma_{B}$ for $B_{L}$).
\label{fig:nearr_all}
}
\end{figure*}

\begin{figure*}
\plotone{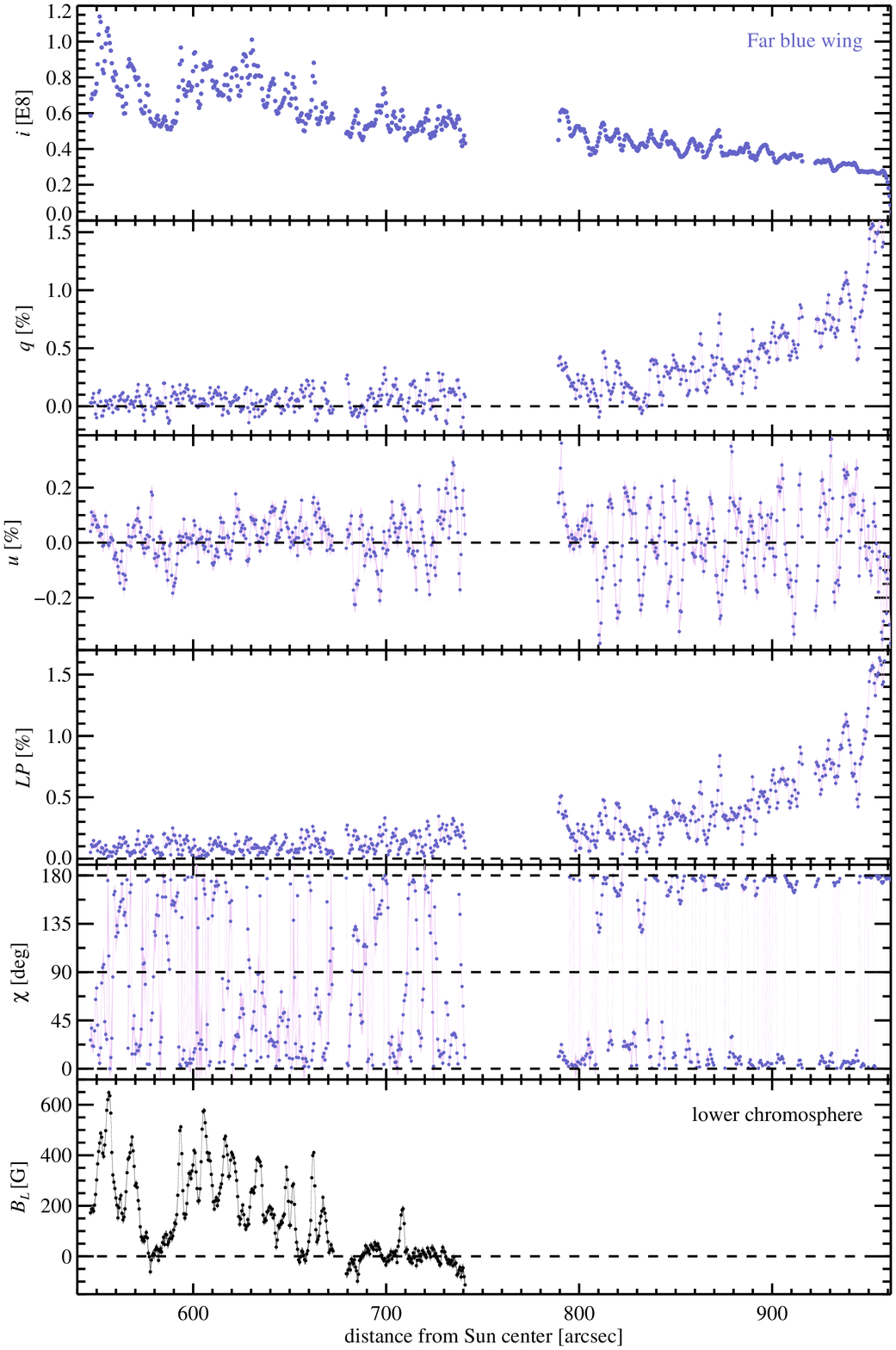}
\caption{From top to bottom: spatial variations of $i$, $q$, $u$, $LP$, and $\chi$ at the far blue wing of the Mg~{\sc ii} $k$ line (the wavelength range indicated in violet color in Fig.~\ref{fig:iquv_spectra}) and the longitudinal magnetic field $B_\mathrm{L}$ at the bottom of the chromosphere derived from the Mn~{\sc i} lines.
The shaded areas show the $\pm1\sigma$ errors ($\pm\sigma_{l}$ for $q$, $u$, and $LP$, $\pm\sigma_{\chi}$ for $\chi$, and $\pm\sigma_{B}$ for $B_{L}$).
\label{fig:farb_all}
}
\end{figure*}

\begin{figure*}
\plotone{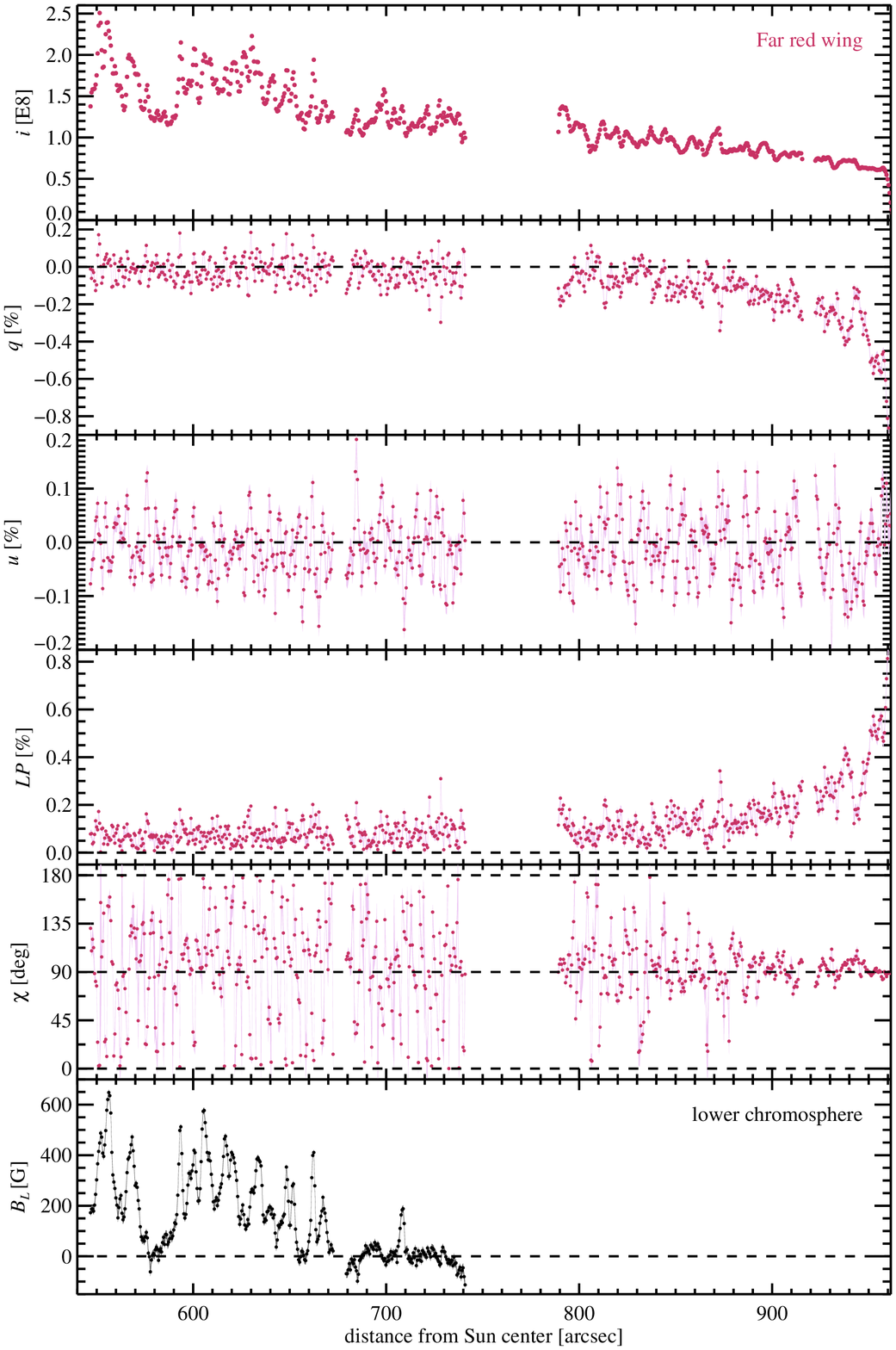}
\caption{From top to bottom: spatial variations of $i$, $q$, $u$, $LP$ and $\chi$ at the far red wing of the Mg~{\sc ii} $k$ line (the wavelength range indicated in magenta color in Fig.~\ref{fig:iquv_spectra}) and the longitudinal magnetic field $B_\mathrm{L}$ at the bottom of the chromosphere derived from the Mn~{\sc i} lines.
The shaded areas show the $\pm1\sigma$ errors ($\pm\sigma_{l}$ for $q$, $u$, and $LP$, $\pm\sigma_{\chi}$ for $\chi$, and $\pm\sigma_{B}$ for $B_{L}$).
\label{fig:farr_all}
}
\end{figure*}

The variation of the quantities characterizing the scattering polarization (i.e., $i$, $q$, $u$, $LP$, and $\chi$ defined in Equations (\ref{eq:i}) - (\ref{eq:chi})) with the distance from the Sun center is shown in Figures~\ref{fig:kcore_all}$-$\ref{fig:farr_all}.
The bottom panels of the figures show the longitudinal components of the magnetic field ($B_{L}$) at each slit position, inferred by applying the WFA to the $V/I$ profiles  
(see Section~\ref{sec:BL}).

\section{Spatial variation of $\lowercase{i}$ and $\lowercase{u}$}
\label{appendix_spatial_iu}
Figures~\ref{fig:i_u_fbw} $-$ \ref{fig:i_u_nrw} represent the spatial variations of $i$ (Equation (\ref{eq:i})) and $u$ (Equation (\ref{eq:u})) at the far blue and red wings and at the near red wing of the $k$ line, which are not shown in the main text.

\section{Hypothesis testing for $u$ zero-crossing}
\label{appendix_u_zero_cross}

In this section, we investigate whether the spatial coincidence of local intensity maxima and $u$ zero-crossings is due to random chance.
For this purpose, we test the hypothesis that the $u$ zero-crossing location is randomly distributed and that local intensity maximum position coincides with $u$ zero-crossing location by chance (i.e., the null hypothesis).

\begin{figure*}
\includegraphics{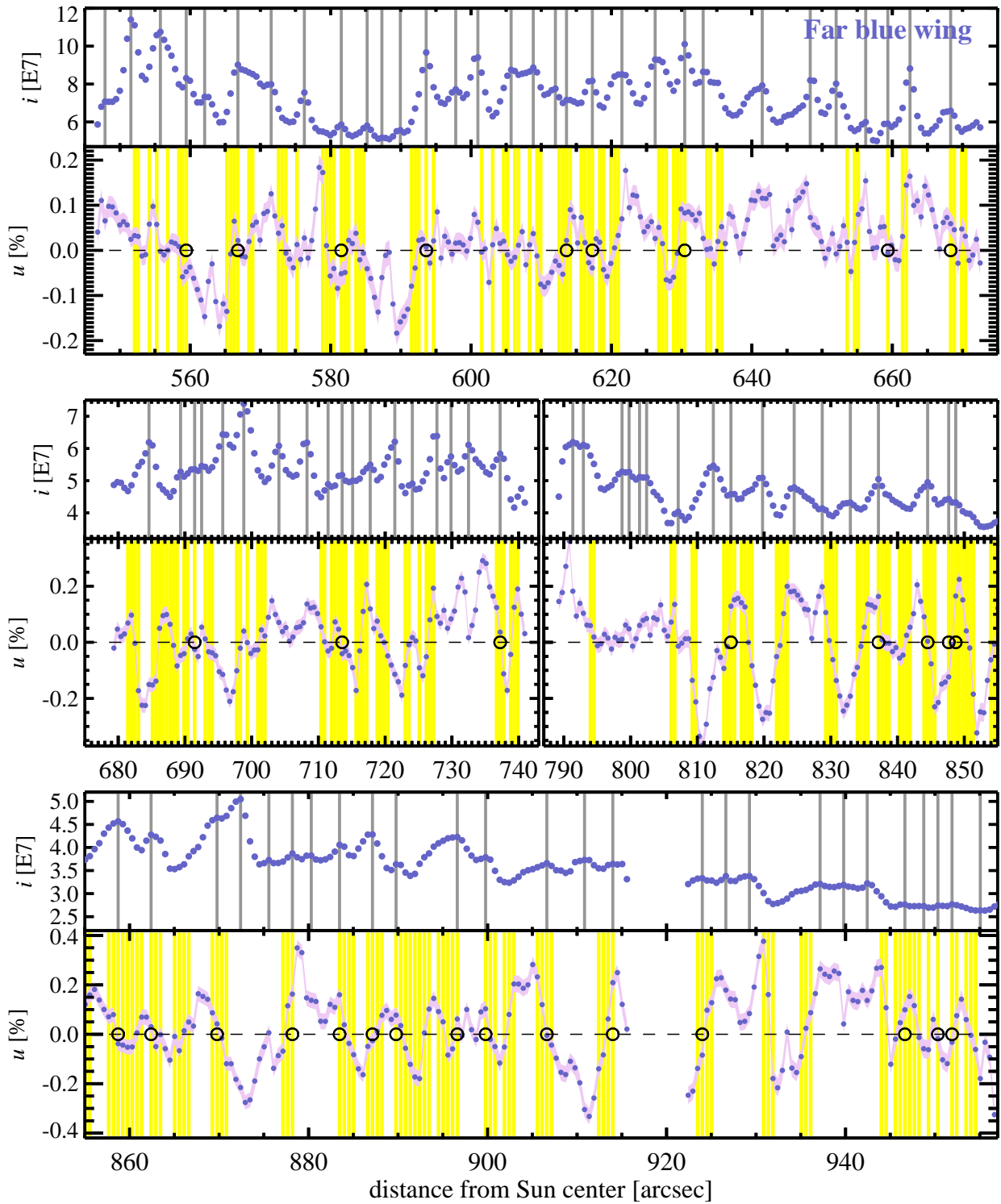}
\caption{
Same as Fig.~\ref{fig:i_u_nbw}, but for the far blue wing of the Mg~{\sc ii} $k$ line (see violet spectral range in Fig.~\ref{fig:iquv_spectra}).
\label{fig:i_u_fbw}
}
\end{figure*}

\begin{figure*}
\includegraphics{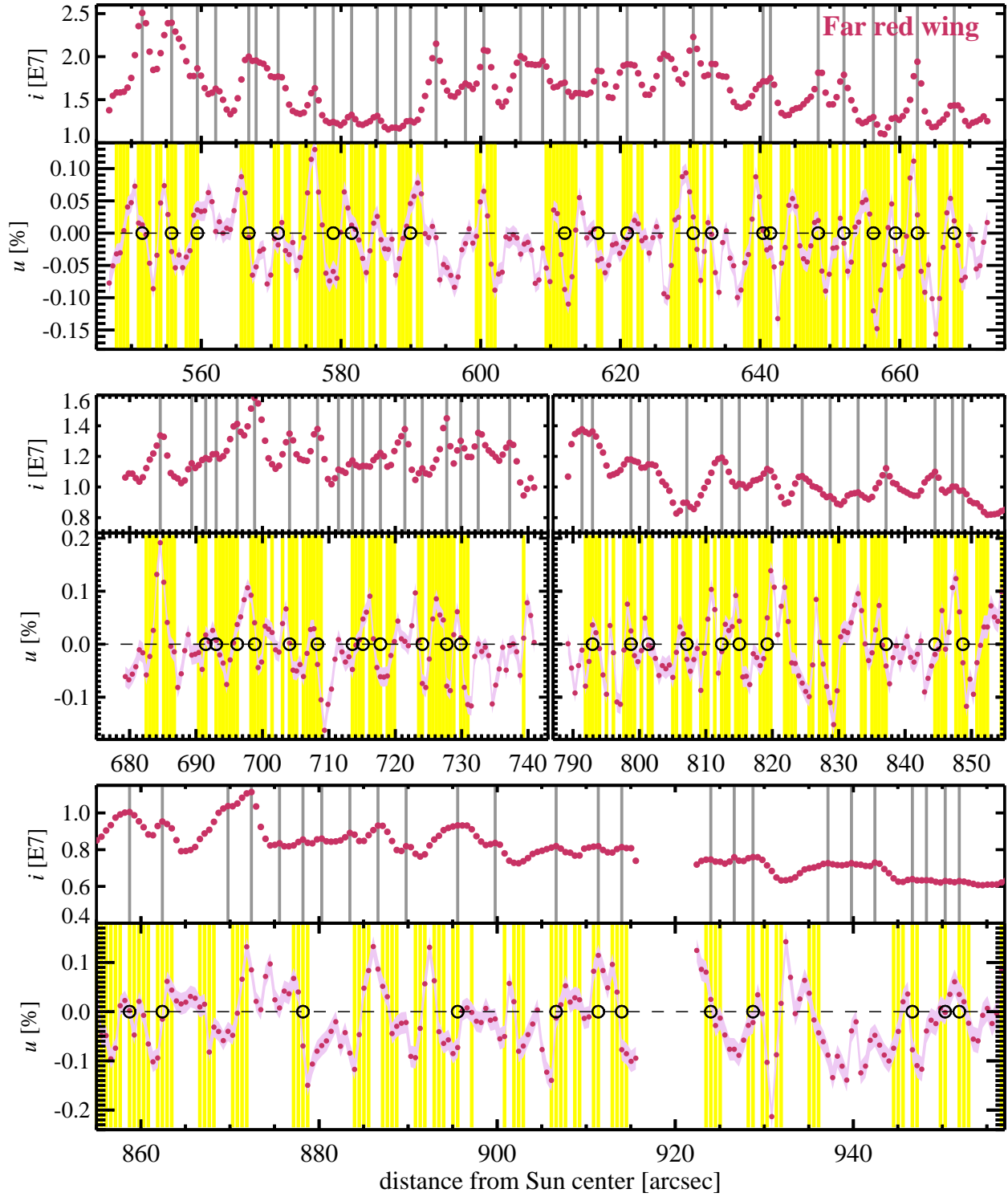}
\caption{
Same as Fig.~\ref{fig:i_u_nbw}, but for the far red wing of the Mg~{\sc ii} $k$ line (see magenta spectral range in Fig.~\ref{fig:iquv_spectra}). 
\label{fig:i_u_frw}
}
\end{figure*}

\begin{figure*}
\includegraphics{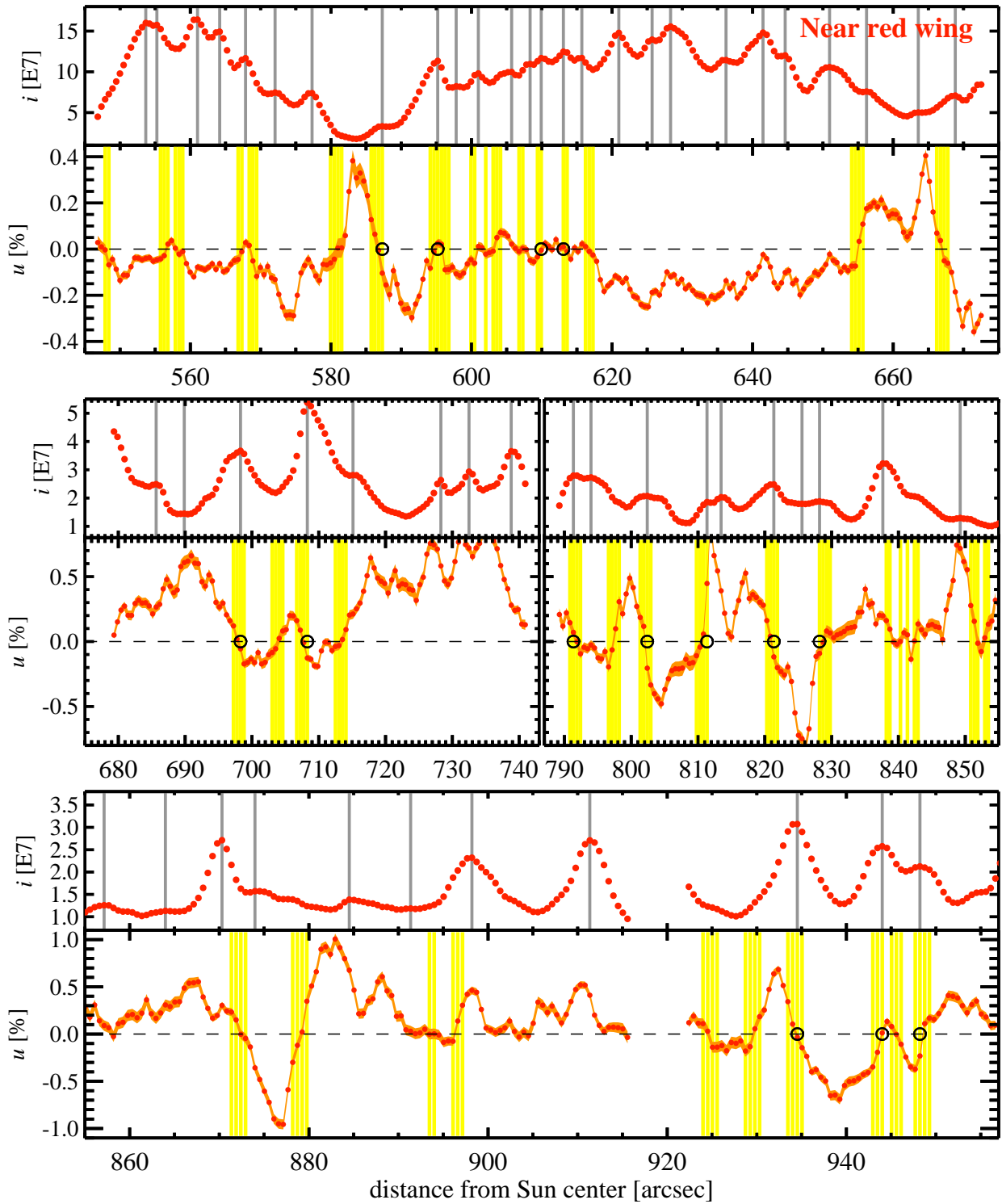}
\caption{
Same as Fig.~\ref{fig:i_u_nbw}, but for the near red wing of the Mg~{\sc ii} $k$ line (see red spectral range in Fig.~\ref{fig:iquv_spectra}).
\label{fig:i_u_nrw}
}
\end{figure*}

Assuming that the probability that a given pixel corresponds to a $u$ zero-crossing location is $p$, the probability that exactly $k$ locations with local intensity maxima are co-spatial with zero-crossing locations out of $n$ locations follows the binomial distribution. The mean and standard deviation of the number of coincidences are $\mu=np$ and $\sigma = \sqrt{np(1-p)}$.
Defining the number of pixels which are identified to be $u$ zero-crossing locations as $s$ (pixels indicated with yellow vertical stripes in Figures~\ref{fig:i_u_nbw}, \ref{fig:i_u_core}, \ref{fig:i_u_fbw}$-$\ref{fig:i_u_nrw}) and the total numbers of pixels in the plage and quiet regions as $S$,
the probability $p$ is given by $s/S$. 

These statistical values, for quiet and plage regions, are tabulated in Tables \ref{tab:iu_statistic_qs} and \ref{tab:iu_statistic_pl}, respectively.
In the quiet region, the identified number of local intensity maxima that are co-spatial with $u$ zero-crossing locations exceeds the number from random chance at the $1\sigma$ level at all considered wavelength ranges, and at $2\sigma$ level in the core and red wing ranges. 
On the other hand, in the plage region, the number of local intensity maxima that is co-spatial with $u$ zero-crossing locations is consistent with random chance within the $1\sigma$ range 
for the core, near red and far blue wings of the $k$ line.
For the near blue wing, the number of spatial coincidences is $1\sigma$ below the mean for the random case and, for the far red wing, the number is $2\sigma$ larger than the mean (as in the quiet region).
Further statistics are needed to reject the null hypothesis confidently at $2\sigma$ level: $n\ge38$ for the near blue wing and $n\ge192$ for the far blue wing in the quiet region, and $n\ge51$ for the near blue wing in the plage region.

\begin{table*}
\begin{center}
\begin{tabular}{ccccccc}
\multicolumn{7}{c}{Quiet Sun ($S=601$ pix)}\\
& \#$i$ max  & \#pix of zero-cross & $p$  & $\mu$ & $\sigma$ & \#$u$ zero-cross \\
& $n$ & $s$ & $s/S$  & $np$ & $\sqrt{np(1-p)}$ &  \\
\hline\hline
Mg~{\sc ii}~$k$ core & 30 & 86 & 0.14 & 4.2 & 1.9 & 10$^{\dagger\dagger}$\\
Near blue wing           & 31 & 101 & 0.17 & 5.2 & 2.1 & 9$^{\dagger}$\\
Near red wing            & 30 & 86 & 0.14 &  4.2 & 1.9 & 10$^{\dagger\dagger}$ \\
Far blue wing           & 62 & 183 & 0.30 & 18.6 & 3.6 & 23$^{\dagger}$\\
Far red wing            & 59 & 210 & 0.34 &  20.1 & 3.6 & 34$^{\dagger\dagger\dagger}$\\
\hline
\end{tabular}
\caption{
Number of $i$ maxima ($n$), number of pixels with $u$ zero-crossings ($s$),
probability of a pixel being a $u$ zero-crossing ($p=s/S$), expected number of $u$ zero-crossing with local $i$ maxima assuming the random distribution ($\mu$) and the standard deviation ($\sigma$), and identified number of $i$ maxima co-located with $u$ zero-crossing (number of $u$ zero-crossing points in Table \ref{tab:iu_peak}) in the quiet region.
In the rightmost column, the number of spatial coincidences between local $i$ maxima and $u$ zero-crossing
that exceeds $\mu+\sigma$, $\mu+2\sigma$, and  $\mu+3\sigma$
are indicated with $\dagger$, $\dagger\dagger$, and $\dagger\dagger\dagger$, respectively.
}
\label{tab:iu_statistic_qs}
\end{center}
\end{table*}

\begin{table*}
\begin{center}
\begin{tabular}{ccccccc}
\multicolumn{7}{c}{Plage ($S=330$ pix)}\\
& \#$i$ max  & \#pix of zero-cross & $p$  & $\mu$ & $\sigma$ & \#$u$ zero-cross \\
& $n$ & $s$ & $s/S$  & $np$ & $\sqrt{np(1-p)}$ &  \\
\hline\hline
Mg~{\sc ii}~$k$ core & 24 & 51 & 0.15 & 3.6 & 1.7 & 5 \\
Near blue wing           & 20 & 24 & 0.07 & 1.5 & 1.2 & 0$^{\ast}$\\
Near red wing            & 26 & 50 & 0.15 &  3.9 & 1.8 & 4 \\
Far blue wing           & 31 & 76 & 0.23 & 7 & 2.3 & 9 \\
Far red wing            & 33 & 125 & 0.38 &  12.5 & 2.8 & 21$^{\dagger\dagger}$\\
\hline
\end{tabular}
\caption{
Same as Table \ref{tab:iu_statistic_qs}, but in the plage region.
In the rightmost column, the numbers of spatial coincidences between local $i$ maxima and $u$ zero-crossing that are smaller than $\mu-\sigma$ are indicated with $\ast$.
}
\label{tab:iu_statistic_pl}
\end{center}
\end{table*}

\bibliographystyle{aasjournal}



\end{document}